\documentclass[manuscript]{aastex}

\slugcomment{To appear in the Astrophysical Journal}

\shorttitle{\emph{Spitzer} Infrared Observations of Hotspots in Radio Lobes}

\shortauthors{Werner et al.}

\begin{document}

\title{\emph{Spitzer} Observations of Hotspots in Radio Lobes}

\author{Michael W. Werner\altaffilmark{1}, David W. Murphy\altaffilmark{1}, John H. Livingston\altaffilmark{1}, Varoujan Gorjian\altaffilmark{1}, Dayton L. Jones\altaffilmark{1}, David L. Meier\altaffilmark{1}, Charles R. Lawrence\altaffilmark{1}, Anthony C. S. Readhead\altaffilmark{2}}

\altaffiltext{}{\it{Copyright 2012 California Institute of Technology. Government sponsorship acknowl-
edged.}}

\altaffiltext{1}{Jet Propulsion Laboratory, 4800 Oak Grove Drive, Pasadena, CA 91109}

\altaffiltext{2}{California Institute of Technology, 1200 East California Boulevard, Pasadena, CA 91125}

\begin{abstract}
We have carried out a systematic search with \emph{Spitzer} Warm Mission and archival data for infrared emission from the hotspots in radio lobes that have been described by Hardcastle et al.\ (2004).  These hotspots have been detected with both radio and X-ray observations, but an observation at an intermediate frequency in the infrared can be critical to distinguish between competing models for particle acceleration and radiation processes in these objects.  Between the archival and warm mission data, we report detections of 18 hotspots; the archival data generally include detections at all four IRAC bands, the Warm Mission data only at 3.6 $\mu m$.

Using a theoretical formalism adopted from Godfrey et al.\ (2009), we fit both archival and warm mission spectral energy distributions [SEDs] -- including radio, X-ray, and optical data from Hardcastle as well as the \emph{Spitzer} data -- with a synchrotron self-Compton [SSC] model, in which the X-rays are produced by Compton scattering of the radio frequency photons by the energetic electrons which radiate them.  With one exception, a SSC model requires that the magnetic field be less or much less than the equipartition value which minimizes total energy and has comparable amounts of energy in the magnetic field and in the energetic particles.  This conclusion agrees with those of comparable recent studies of hotspots, and with the analysis presented by Hardcastle et al.\ (2004). We also show that the infrared data rule out the simplest synchrotron only models for the SEDs.  We briefly discuss the implications of these results and of alternate interpretations of the data.
\end{abstract}


\section{Introduction}

Hotspots in the extended lobes of powerful (FR II) radio galaxies have been known since the early work on Cygnus A (Mitton \& Ryle 1969; Miley \& Wade 1971; Hargrave \& Ryle 1974).  The hotspots arise at the interface where a jet accelerated by a black hole at the center of the galaxy plows into the local intergalactic medium. Shocks created at this interface accelerate electrons to relativistic energies, and synchrotron radiation from the electrons produces both the radio hotspot and, as the electrons diffuse away from the hotspot, the extended emission from the lobe. Determining the physical parameters of the hotspots is important both for understanding radio galaxy energetics and as a way of refining models of particle acceleration and physical conditions in astrophysical plasmas.  Reviews of the properties of hotspots and the problems they pose are provided by Scheuer (1982), Carilli et al.\ (1991), Hardcastle et al.\ (2007), and O'Dea et al.\ (2009).

A major step forward in our quantitative understanding of this phenomenon resulted from the X-ray detection of hotspots, vastly increasing the frequency baseline over which they are observed.   This was done in the first instance by ROSAT observations of Cygnus A, but a thorough analysis of \emph{Chandra} results (Hardcastle et al.\ (2004), hereinafter H04; Massaro et al.\ (2010)) led to X-ray detections of 43 hotspots associated with 3C radio galaxies. (Hardcastle et al also report upper limits on the X-ray emission from another 22 hotspots; these are not systematically discussed in the present paper.) These data are presented and analyzed by H04, together with radio frequency (1.5, 4.8, or 8.5 GHz) data on the hotspots.   H04 suggest that there are two classes of hotspots. The archetype of one class is Cyg A hotspot A, which was shown by Harris, Carilli, and Perley (1994) to exhibit a broadband spectral energy distribution (SED) which is not consistent with a simple synchrotron emission model, but an SSC model with a magnetic field close to equipartition could explain the radio and X-ray emission. This was confirmed with infrared observations by Stawarz et al.\ (2007), who detect the hotspot with Spitzer's IRAC camera and show that its SED -- including infrared measurements and optical upper limits as well as radio and X-ray data -- can be interpreted as follows:  A sharp spectral steepening at around $10^{13}$ Hz is attributed to a break in the power law distribution of the relativistic electrons, and the flux in the X-ray region is attributed to Compton scattering of the radio photons by the energetic electrons which produce the radio emission.  This synchrotron self-Compton [SSC] model provides a good fit to the overall SED and shows that the equipartition condition of comparable particle and field energies applies -- within a factor of a few -- to this hotspot.  Godfrey et al.\ (2009) show that the radio to X-ray SED of a hotspot in PKS 1421-490 is also well fit by a SSC model which, again, is close to equipartition.

By contrast, H04 show that for many hotspots the observed X-ray flux is much higher than that predicted from a SSC model pegged to the radio data and assuming equipartition.  A simple, but not unique, interpretation is that in these hotspots the X-rays produced by the SSC process are negligible compared to an extension of the underlying synchrotron emission spectrum into the X-ray.  H04 point out that hotspots with X-ray emission consistent with the SSC model are systematically more luminous than others, perhaps because the high radio frequency photon density in these sources makes it difficult to accelerate electrons to the energies required to produce X-rays by synchrotron emission.   X-ray and radio data alone cannot always distinguish between the SSC and synchrotron only interpretations of the spectral energy distributions.  As was the case for Cyg A and for PKS 1421-490, a visual or infrared measurement that lies at frequencies midway between the radio and the X-ray may be critical to making this distinction firmly.  

The \emph{Spitzer} \emph{Space Telescope}, with the superb sensitivity of its IRAC camera at wavelengths from 3.6 to 8 $\mu m$, allows us to augment the existing data with infrared observations of hotspots at frequencies intermediate between the X-ray and the radio. In addition to the work of Stawarz et al.\ (2007), Tingay et al.\ (2008) and Kraft et al.\ (2007) have reported Spitzer results on hotspots in Pic A and 3C33, but there has been no report of a systematic study of hotspots with Spitzer. For that reason, working with the results reported by H04, we have carried out with warm \emph{Spitzer} 3.6 $\mu m$ observations of 25 of the X-ray detected hotspots, and we have also analyzed archival data on 9 hotspots previously observed by \emph{Spitzer}. The archival data generally were taken in all four IRAC bands at 3.6, 4.5, 5.8, and 8 $\mu m$.  We analyze in detail here data on 17 hotspots that are robustly detected at radio, infrared, and X-ray wavelengths.  We also analyze the data on 7 other hotspots for which there are X-ray and radio detections but only upper limit[s] in the infrared.   When available, we include in the analysis optical upper limits or detections from H04.  We compare the results with the predictions of simple SSC models, following the formalism of Godfrey et al.\ (2009) and discuss other possible interpretations of the data.

\section{Observations and Data Reduction}

\subsection{Target Selection and Observational Strategy}

For each hotspot with radio and X-ray detections, we calculated the 3.6 $\mu m$ flux of a simple power law connecting the X-ray and 5 GHz radio data.  The 3.6 $\mu m$ flux predicted in this way, which we call 3.6p, lies between $\sim$0.5 and 50 $\mu Jy$.  For those sources noted by H04 to be brighter in the X-ray than the predictions of the SSC model, we assumed that the true SED is in fact such a power law, and selected the 3.6 $\mu m$ integration time to measure 3.6p with S/N=10.  For the sources that appeared consistent with the SSC model, we set the integration time to achieve S/N = 3 on a flux level equal to 0.25*3.6p.  This approach is predicated on the observation that the measured 3.6 $\mu m$ flux of the Cyg A hotspots is about 25\% of the value of 3.6p derived using the above interpolation.

After establishing this criterion, we examined the \emph{Spitzer} archive and found that nine of the hotspots had already been observed with integration times adequate to reach within a factor of two of our S/N criterion.   We deleted these from our observing list in favor of using the archived data, which was generally available at all four IRAC bands; in the case of Pic A W we also use an archival 24 $\mu m$ detection.  Both newly observed and archival hotspots included in this program are listed in Table 1, together with the angular size and redshift given by H04 and the radio position derived as described below.  Note that some of these hotspots have only X-ray upper limits but are tabulated by H04 and serendipitously lie in the IRAC frame. In the data presentation we separate the archival sources, generally detected in all four IRAC bands, from the newly observed hotspots.  Neither sample can be considered to be unbiased, and they appear to differ from one another:  although the two samples have similar median radio fluxes, we note that the X-ray and optical brightnesses are generally greater for the archival sample.  Nevertheless, the principal conclusions of this work are the same for both samples. 

\subsection{Identifying a Hotspot}
Many sources appeared in the \emph{Spitzer} images, so care was required to identify those associated with hotspots.  To do this, we determined the position of each radio hotspot by careful measurement -- albeit by eye and hand -- of the highest frequency radio image available in the literature.  The adopted positions are given in Table 1.  These positions have a nominal uncertainty of $\sim1$".  We then downloaded the X-ray image for each hotspot from the \emph{Chandra} archive and overlaid the infrared and X-ray images to verify that the emission at these two bands was in fact coincident with the radio hotspot.  Because the \emph{Spitzer} and \emph{Chandra} images have positional uncertainties $<1"$, we could verify that sources detected in all three bands -- radio, X-ray, and infrared -- are cospatial to within better than 2 arcsec with one possible exception, 3C 173.1 N.  The same procedure was used with the archival data, facilitated in some cases by the fact that the hotspots are generally brighter at 8 than at 3.6 $\mu m$, while the density of potentially confusing sources is lower. Existing data sets, including both the Spitzer and the Chandra images described here and the archival radio images, might support a more detailed comparison of the source positions in different spectral bands.  Such an astrometric exploration was beyond the scope of the present study, so we have assumed unless otherwise stated that the images in the X-ray, infrared and radio bands coincide precisely and that all observed radiation arises from the same homogeneous spatial volume. Figures 1 through 9 show the X-ray and infrared images of several hotspots with the radio position from Table 1 overlaid according to the coordinate system established in the FITS header of the images.

\subsection{Photometry and Calibration}
We planned the observations with appropriate integration times and dither strategy to meet our pre-established criteria.  The observations were executed by \emph{Spitzer} during the 08/11/2009 to 03/13/2010 time period and reduced with the warm mission pipeline.  We used appropriate standard pipeline products both for the newly-executed observations and for the archival images that we analyzed.  Overall, we detected 8 hotspots in the archival data and 10 -- out of a total of 18 that could be cleanly observed -- in the new warm mission data. One of the warm mission infrared detections, 3C 173.1 N, has only an X-ray upper limit

We carried out the photometry by measuring the flux into a 2.4" (2 pixel SCM, 4 pixel SWM) radius aperture centered on the radio position from Table 1.  A correction for the sky brightness was obtained by using the median brightness of a reference annulus centered on this position with inner radius 14.4 arcsec and outer radius 24 arcsec.  Finally, an aperture correction was applied following the prescription provide by the Spitzer Science Center.  (For Pic A at 24 $\mu m$, the photometry aperture had a radius of 3.5 arcsec, and the sky annulus radii were 20 and 32 arcsec). 

The results of the warm mission observations are given in Table 2, and the 9 hotspots for which we relied entirely on archival data are listed in Table 3 (we exclude the four Cygnus A hotspots for which the IRAC data has been published by Stawarz et al.\ (2007)).  Note that all detected archival hotspots are detected in all four bands, with the exception of 3C 275.1 N which was too close to the edge of the mosaic for reliable measurements at 3.6 and 5.8 $\mu m$.  The upper limits are 3-sigma; the tabulated uncertainties are derived from the noise image provided by the SSC and do not include several systematic effects described below.

It is noteworthy that the 3.6 to 8 $\mu m$ SEDs for the archival hotspots (Table 3) show flux density increasing with wavelength, a spectral slope consistent with that expected for non-thermal emission. We investigated whether it would be necessary to apply a color correction to the measured fluxes, because the expected shape of the infrared spectrum of a hotspot differs from that of the stellar spectra used to establish the IRAC flux scale.  However, documentation available from the \emph{Spitzer} Science Center website http://ssc.spitzer.caltech.edu/irac/iracinstrumenthandbook/21/ shows that this correction is no bigger than 1.5\% at any observed wavelength even for spectra as steep as a power law with flux density falling as $\nu^{-1.5}$, which is about the median slope for the infrared SEDs we observe.  Therefore we did not apply a color correction to any of our sources, and the flux calibration given as part of the standard  {\it Spitzer} Science Center data products was adopted for this work.

\section{Results}
\subsection{Detections, Upper Limits, and Uncertainties}
In the subsequent sections of this paper, we analyze in detail the data on the 17 hotspots in Tables 2 and 3 that are robustly detected in all three bands -- radio, infrared, and X-ray.  The optical data are used in the analysis as well.  Tables 2 and 3 also contain several sources with X-ray and radio detections but only an upper limit in the infrared. The upper limit to the infrared flux density associated with an undetected hotspot is determined by estimating the noise within a sky region equivalent to our photometry aperture and taking three times that value as our upper limit. This was accomplished by fitting a Gaussian to a histogram of the pixel values of the entire image and taking the GaussianÕs half width at half maximum to be the sigma per pixel for the image. Then for the number of pixels in our photometry aperture we added the sigmas per pixel in quadrature to determine the noise in the aperture and finally we multiplied by three to derive our three sigma upper limit. We analyze the data on the seven hotspots with X-ray and radio detections and infrared upper limits to show that the non-detections do not imply a physically different picture from that suggested by the detections.

There are several potential sources of uncertainty in the measurements in addition to the largely statistical ones tabulated in Tables 2 and 3 above. Firstly, there is a calibration uncertainty of $\pm3$\%. More significantly, there is a systematic uncertainty due to the possible inclusion in the measurement beam of radiation from the portions of the radio lobes contiguous to the hotspot. To estimate the magnitude of this effect we make use of the fact that the PicA W hotspot, our nearest target at z $\sim$ 0.035, is clearly resolved in all five bands (see Figures 6--8).  The flux into a $\sim$15" aperture which encompasses the extended emission is typically 40\% greater than that attributed to the hotspot using the measurement protocol described earlier, and tabulated in Table 3.  Given that for a more distant target the contribution from the lobes would be unresolved, we feel that the fluxes tabulated in Tables 2 and 3 could overestimate the hotspot flux by up to 40\%.  As discussed below in Section 4.6, a reduction in hotspot fluxes to account for this factor does not qualitatively change the conclusions of this paper.

Finally, at the suggestion of the referee we have investigated the possible effects of source confusion on these results.  Taking the median 3.6 $\mu m$ flux density of 15 $\mu Jy$ for the warm mission sources as a point of comparison, we see from Fazio et al.\ (2004) that there are approximately $1.5\times10^{-4}$ sources per square degree brighter than this at 3.6 $\mu m$ -- predominantly galaxies -- at moderate to high galactic latitude.  The effective \emph{Spitzer} beam for this comparison is no bigger than our 2.4 arcsec photometric aperture, which leads to a confusion probability of order 0.5\% for any particular object.  Thus there is a $<$10\% chance that one of our warm mission detections results from a chance superposition of a background galaxy.  For the archival sources, the chance of confusion is negligible, considering that they are all brighter than $\sim25$ $\mu Jy$ at 3.6 $\mu m$ and also that they are all detected at 8 $\mu m$ with flux levels considerably greater than at 3.6 $\mu m$ which is the expectation from a hotspot SED but not a normal galaxy SED.  We conclude that the results presented here are not seriously affected by confusion.

\subsection{Comments on Individual Sources}
\subsubsection{3C 173.1 N}
Perlman et al.\ (2010) present high resolution radio, X-ray, and infrared images of hotspot 3C 445 S, not included in our sample, and call attention to the fact that the image centroids in the three bands are spread over some 2 to 4 arcsec.  The infrared data in this case were taken at J, H, and K with the VLT and have considerably higher resolution than our \emph{Spitzer} data.  They present a detailed model, which explains both the displacement and the observed SED of the hotspot.  As mentioned above, the X-ray, radio, and infrared images of the hotspots we have studied generally coincide to within $\sim$1".  The only source for which this definitely appears not to be the case is 3C 173.1 N, for which the infrared position is $\sim$2" west of the radio position, in a direction transverse to the jet.  However, this hotspot was not detected in the X-ray and is thus not included in the analysis presented here.

\subsubsection{Pic A W}
Pic A W, previously studied with Spitzer by Tingay et al.\ (2008) is the only hotspot in our sample which is clearly resolved by Spitzer (see Figures 6-to-8).  Emission is seen not only from the hotspot but also from the bright outer regions of the adjacent radio lobe.  Our flux measurements of the hotspot (Table 3), made on the same archival images used by Tingay et al.\, are systematically (on average) 8\% lower than theirs, which is attributable to the fact that we used a smaller aperture.  Pic A W is also the only hotspot for which [archival] 24 $\mu m$ data are available.

\section{Modeling and Discussion}
\subsection{Introduction}
There are several competing emission mechanisms that can potentially explain the observed hotspot radio-to-X-ray Spectral Energy Distributions (SEDs) including: Synchrotron Self-Compton (SSC) with the IC seed photons being internally produced by the synchrotron emission, Synchrotron External-Compton (SEC) where the majority of the IC seed photons come from an external source such as the Cosmic Microwave Background (CMB) or some other external source and Synchrotron Only (SO) with no significant Inverse Compton (IC) emission for the frequency range under study. In order to attempt to distinguish between these different emission mechanisms we begin by modeling our hotspot SEDs using a SSC model since at some level SSC emission must be occurring and it is a useful straw-man emission mechanism since it lies between the SO and SEC emission mechanisms which have no significant or enhanced IC emission respectively. We also examine a new variant of the SO model which we call the SO Compact Diffuse (SO-CD) model in which each hotspot is composed of two components, a compact and a more diffuse component.

\subsection{SSC Modeling Introduction}
We use the single-zone spherical SSC model developed by Godfrey et al.\ (2009) [hereafter referred to as G09] to model the multi-wavelength emission of 24 hotspots we studied for which we obtained either IR measurements (17 hotspots) or IR upper limits (7 hotspots). These hotspots were all studied by H04 and have both X-ray and radio detections. We divide them into our new observations from the \emph{Spitzer} Warm Mission (SWM) or our reprocessing of archival data from the \emph{Spitzer} Cold Mission (SCM) (see Table 2 and Table 3). For the SWM hotspots we have 9 IR detections and 6 IR upper limits and for the SCM hotspots we have 8 IR detections and 1 IR upper limit. For our SSC modeling we used the 5 GHz radio flux densities ($S_{R}$), 1 keV X-ray flux densities ($S_{X}$), redshift (z), optical fluxes or upper limits [almost entirely from \emph{HST}, and hotspot angular sizes ($\theta_{H}$) as listed in Table 3 of H04 (which are also reproduced in our Table 2 and Table 3). We use $S_{IR}$  and  $S_{O}$ to  refer to the 3.6 $\mu m$ IR and 0.55-0.70 $\mu m$ optical flux densities (except for 3C 275.1 N which has no 3.6 $\mu m$ flux density so a 4.5 $\mu m$ flux density will be used in its place).  Corresponding to the $S_{R}$, $S_{IR}$, $S_{O}$, and $S_{X}$  flux densities we will use the  nomenclature $\nu_{R}, \nu_{IR}, \nu_{O}, \nu_{X}$ and $SE_{R}, SE_{IR}, SE_{O}, SE_{X}$ for their respective frequencies and spectral energies, where $SE =$ Spectral Energy $= \nu S_{\nu}$ for a  flux density $S_{\nu}$ at observed frequency $\nu$.

For those hotspots listed as cylinders in H04 we converted the cylinder parameters to radii of spheres having the same volume. One other parameter from H04's Table 3 that we also used as a comparison with our work was the R ratio, which we call $R_{H04}$, which is the ratio of the observed X-ray flux density to the X-ray flux density if the hotspot were in equipartition as determined by H04. One of the major findings of the H04 paper is that if the hotspot emission mechanism is SSC, then the typical hotspot has a magnetic field strength well below its equipartition value and consequently has a $R_{H04}$  value  much greater than unity. For these $R_{H04} \gg 1$ hotspots (which are at lower redshifts and radio luminosities than their $R_{H04} \approx 1$ counterparts) H04 advocate the SO model, since that model alleviated the need for the hotspots to be so far out of equipartition and was consistent with their optical detections and upper limits.   

However, because we have actual detections of, or stringent upper limits on, the hotspot infrared flux from \emph{Spitzer}, we are able to severely constrain the SO model and rule it out for many of our observed hotspots, in favor of the SSC or other models.  This will lead to robust determinations of whether hotspots are in equipartition or not.  

The SSC model requires two co-located physical components: a magnetic field (of strength B) and relativistic electrons. In our modeling, like H04, we assume that relativistic electrons are the only significant component to the particle energy density (i.e., like H04 we use $\kappa = 0$ in our equivalent of H04's equation (B2) where $\kappa$ is the ratio of energy densities of non-radiating to radiating particles). We also use the G09 model for the relativistic electron Lorentz gamma ($\gamma$) distribution $dN/d\gamma$, which is the number of relativistic electrons per unit volume per unit Lorentz factor which is given by their equations (3) and (4) which we re-state as::
\begin{equation}
\frac{dN}{d\gamma} = K_e\gamma_b\gamma^{-(p+1)}\frac{1-(1-(\gamma/\gamma_b))^{p-1}}{p-1}
\end{equation}
for $\gamma_{1} < \gamma < \gamma_{b}$, and
\begin{equation}
\frac{dN}{d\gamma} = K_e\gamma_b\frac{\gamma^{-(p+1)}}{p-1}
\end{equation}
for $\gamma_{b} < \gamma < \gamma_{2}$, where $K_{e}$ is a constant for a particular hotspot, and $\gamma_{1}, \gamma_{b}$, and $\gamma_{2}$ are the minimum, break, and maximum relativistic electron Lorentz factors respectively. This functional form for $dN/d\gamma$ produces a broken power law spectrum in which the spectral index changes smoothly from $-p$ to $-(p+1)$ at about $\gamma = \gamma_{b}$. Also, for $\gamma \ll \gamma_{b}$, equation (1) simply becomes $dN/d\gamma = K_e\gamma^{-p}$ to a very good approximation. We adopt the same symbols as used by G09 except for the $\gamma_{1} < \gamma < \gamma_{b}$ power-law index $-p$ which is used by H04 and is also more generally used in the literature than G09's $-a$ for this index. The break in the relativistic electron distribution at   $\gamma = \gamma_{b}$ is due to synchrotron spectral aging and can be used to estimate how long the electrons have stayed in the hotspot region. Like many authors, instead of fitting flux densities we prefer to fit the corresponding $SEs$. From basic synchrotron theory, to a good approximation, the $SE$ spectral index $\alpha_{SE} (= -d(\ln(SE))/d(\ln(\nu)) )$ is approximately given as follows: $\alpha_{SE} = (p-3)/2$ for  $\gamma_{1} < \gamma < \gamma_{b}$ and $\alpha_{SE} = (p-2)/2$ for  $\gamma_{b} < \gamma < \gamma_{2}$.  Also, in general, the flux density spectral index $\alpha_{\nu} (= -d(\ln(S_{\nu}))/d(\ln(\nu))$ is related to $\alpha_{SE}$ by the simple relationship: $\alpha_{SE} = \alpha_{\nu}-1$. In all our fitting we adopt the same redshift, hotspot angular size, and the values $\gamma_{1}$ and $p$ (which were 1000 and 2 respectively) used by H04. With $p=2$, $\alpha_{SE} = -0.5$ and $\alpha_{SE} = 0$ for  $\gamma_{1} < \gamma < \gamma_{b}$ and   $\gamma_{b} < \gamma < \gamma_{2}$ respectively. $p=2$ is a good physical assumption that can be produced by first-order Fermi processes as electrons cross and re-cross the hotspot terminal shock at the end of the jet (Bell 1978).

For the model-fitting we also used a cosmology in which $H_{0} = 72  km s^{-1} Mpc^{-1}, \Omega_{m} = 0.26$ and $\Omega_{V} = 0.74$ which differ very slightly from the values used by H04 which were $70 km s^{-1} Mpc^{-1}$, 0.3 and 0.7 respectively.

With the synchrotron emission due to relativistic electrons at  $\gamma_{1}, \gamma_{b}$ and $\gamma_{2}$, we may associate characteristic frequencies in the observerÕs frame $\nu_{1}\leq\nu_{b}\leq\nu_{2}$ given by:
\begin{equation}
\nu_{1,b,2} = \left(\frac{\delta_H}{1+z}\right)\left(\frac{3}{4\pi}\right)\left(\frac{\Omega_0}{B}\right)B\gamma_{1,b,2}^{2}
\end{equation}
where $\delta_H =$ hotspot Doppler factor, $z =$ source redshift, $\Omega_0 =$ nonrelativistic gyrofrequency, and $B = $hotspot magnetic field flux density.

The $\delta_H/(1+z)$ factor converts the frequency from the hotspot-emitted frame to the observerÕs frame in which we do all our model-fitting and $\Omega_0/B$ is a constant. Using typical values from our model fitting we find that representative values of $\nu_1$ and $\nu_2$ are given by:
\begin{equation}
\nu_1 \approx 42\frac{\delta_H}{1+z}\frac{B}{10 \mu G}\left(\frac{\gamma_1}{1\times10^{3}}\right)^{2} {\rm ~MHz,  ~and~} \\
 \nu_2 \approx 42\frac{\delta_H}{1+z}\frac{B}{10 \mu G}\left(\frac{\gamma_2}{1\times10^{6}}\right)^{2} {\rm ~THz}
\end{equation}

Like H04, we assume $\gamma_{1} = 1000$ and with this assumption $\nu_{1}$ is always well below 5 GHz, which simplifies the SSC modeling. Other authors have modeled particular hotspots in more detail from their low radio frequency emission and indeed find $\gamma_{1} \approx 1000$ for these hotspots. For example, G09 find $\gamma_{1} = 650$ for the northern hotspot in PKS 1421-490 assuming $\delta_{H}=1$. For this hotspot, $\nu_{2}$ corresponds to a wavelength of $\approx 10 \mu m$  and lies in the infrared. In our fitting, like H04, we assume that any Doppler beaming in the hotspots is negligible and hence we set $\delta_{H} = 1$. 

One important physical question to be determined is if the hotspots are in equipartition or not. We examine this question by calculating what we call the Energy Density Ratio (EDR) which is the ratio of magnetic energy ($U_{B}$) to relativistic electron energy density ($U_{E}$) i.e., $EDR = U_{B}/U_{E}$. EDR = 1 defines the equipartition condition. $U_{B}$ is simply given by $B^{2}/2\mu_{0}$ [in SI Units], or $B^{2}/8\pi$ [in cgs units]. $U_{E}$ is obtained by integrating $(1+ \kappa)(\gamma-1) m_{e}c^{2} (dN/d\gamma)$ from $\gamma_{1}$ to $\gamma_{2}$, but like H04 we set $\kappa = 0$ as mentioned above.  Furthermore, the total relativistic electron density, $N_{e}$, is given by simply integrating $dN/d\gamma$ from $\gamma_{1}$ to $\gamma_{2}$, and so to a good approximation $N_{e} \approx K_{e}/\gamma_{1}$ for $p=2$.

Some authors, such as H04, prefer to work with the ratio $B/B_{eq}$ instead of EDR where $B$ is the fitted magnetic field and $B_{eq}$ is the magnetic field if the hotspot was in equipartition. For $p=2$, which both we and H04 use,  $B/B_{eq} \approx EDR^{2/7}$. This approximation which we will be using to determine $B/B_{eq}$ is such that for typical hotspot parameters (which we will determine in the following subsections) $B/B_{eq}$ can be determined to better than 5\% which is more than adequate for our purposes.  Furthermore, as H04 point out, a key factor in IC emission is that the IC emission is directly proportional to $N_{e}$. For $p=2$, $N_{e}$ is related to its equipartition value $N_{eq}$ by  
\begin{equation}
N_{e}/N_{eq} =  R_{H04} = \left(\frac{B}{B_{eq}}\right)^{-3/2} \approx EDR^{-3/7}
\end{equation}
Thus, if a hotspot has a magnetic field strength 100 times lower than its equipartition value then its IC emission will be 1000 times stronger than the equipartition value and its $R_{H04}$ value will also be equal to 1000.  H04 found using their SSC model that the majority of hotspots were well out of equipartition with $B/B_{eq} \ll 1$ and consequently have $N_{e}/N_{eq} \gg1$ and produce large amounts of IC X-ray emission if the SSC model is correct. This idea seemed somewhat untenable so they considered the alternative SO model with synchrotron emission all the way from the radio to the X-ray.  At the time this seemed an attractive alternative to the SSC model.  

\subsection{Pictor-A W Model-Fitting Example}
As an initial example of our SSC model-fitting results (using the full G09 SSC model) we show in Figure 10 the results of the SSC model-fitting undertaken on the well-studied Pictor A (Pic A) W hotspot, the only hotspot in our sample for which 24 micron data are available. All the subsequent model-fitting described in this paper also uses the full G09 SSC model. On the left of Figure 10 we show the results of what we call 3-fit model-fitting in which we use only the 3 spectral energies $SE_{R}$, $SE_{IR}$, and $SE_{X}$ to determine  the SSC model parameters $B$ and $\gamma_{2}$ as well as the derived parameter $N_{e}$ (which is a function of $K_{e}, \gamma_{1}$, and $\gamma_{2}$).  In the 3-fit model-fitting $\gamma_{b}$ is set  equal to $\gamma_{2}$  since there are not enough $SEs$ fitted to determine both $\gamma_{b}$ and $\gamma_{2}$. In contrast, on the right of Figure 10, we show the results of what we call all-fit model-fitting in which all the measured spectral energies, including optical and all available infrared bands, are used to determine not only $B, \gamma_{2}$, and $N_{e}$, but also $\gamma_{b}$. More details on 3-fit model fitting can be found in Section 4.4 and more details on all-fit model fitting can be found in Section 4.5. We use two fitting procedures because the SWM hotspots are detected only at 3.6 $\mu m$, while the SCM hotspots are observed in all four IRAC bands and frequently detected in the visible as well. As can be seen in the 3-fit model-fitting when the only IR and optical data point is that at  3.6 $\mu m$ there is a poor fit to the other IR and optical data. However, by using all these points and also fitting for $\gamma_{b}$, a much better overall fit is obtained and the peak in both synchrotron  and IC curves are also reduced. For both of these fitting methods the magnetic field strength is well below its equipartition value (3.77 $\mu$G), with $B/B_{eq} \approx 0.020$ and $EDR \approx 1.1 \times 10^{-6}$. However, for this particular bright and well-studied hotspot, there are some difficulties interpreting the X-ray emission as being due to IC emission. Most notably the radio spectral index $\alpha_{R}$ is $0.740\pm0.015$ while the X-ray spectral index is $\alpha_{X} = 1.07\pm0.11$ (Wilson et al.\ (2001)); in a simple SSC model the two slopes should be identical.  This inconsistency shows that for at least this hotspot something else beyond simple SSC emission is occurring. VLBI observations by Tingay et al.\ (2008) show that the Pictor A W hotspot has compact VLBI-scale components, which could produce X-ray synchrotron emission at the right level. This is what we call the SO-CD model. In our discussion section and in Appendix A we examine the SO-CD model in more detail.

\subsection{SSC 3-Fit Model-Fitting}
For the all the hotspots in our sample, we determined SSC model parameters using the 3 $SEs$: $SE_{R}, SE_{IR}$, and $SE_{X}$, and we call this 3-fit model fitting. For those hotspots with no IR detections we used the 3.6 $\mu m$ upper limits for $SE_{IR}$. We did not use the optical flux density detections or upper limits in this fitting. Since we have only 3 SEs availaable for 3-fit model fitting we are unable to solve for $K_{e}, B, \gamma_{b},$ and $\gamma_{2}$.  Consequently, for the 3-fit modelling we set $\gamma_{b} = \gamma_{2}$ to eliminate the need to solve for all four parameters.

 	We developed a non-linear least squares fitting routine based on our version of the G09 SSC model that minimized the fractional $SE$ residual to determine the fitted parameters $B$ and $\gamma_{2}$.  Table 4 shows our fitting results for $B$ and $\gamma_{2}$ as well as the derived parameters $N_{e}, EDR,$ and $B/B_{eq}$. Figures 11, 12, and 13 show plots of the fitted SED synchrotron (in red) and IC SED functions (in blue) for the SWM hotspots with measured IR flux densities, SCM hotspots with measured IR flux densities, and hotspots with only upper limits to their IR flux densities respectively. As can be seen for the red synchrotron curve to a good approximation for $\nu_{1} < \nu \ll \nu_{2}$, $\alpha_{SED} = -0.5$ as is expected for the $p=2$ SSC model we are using and this curve passes through $SE_{R}$ while $SE_{IR}$ constrains the value of $\gamma_2$. 
	In order for the blue IC curve to pass through $SE_{X}$, with the exception of 3C280 W, the magnetic field has to lie below its equipartition value, in some cases by as much as two orders of magnitude. For the SWM hotspots detected in the infrared, Figure 11 shows that the fits are consistent with the optical upper limits when available. Inspection of the SED plots in Figure 11 also shows (cf. Section 5.1) that the SO model can be ruled out in the 7 hotspots with optical upper limits but not the two hotspots with no optical upper limits (i.e., with only 3 measured flux densities: 3C 321 E and 3C452 W). More IR or optical data are needed to distinguish between the two models.

 As can be seen in Figure 12, for the SCM mission hotspots with measured IR flux densities the 3-fit fitted synchrotron curve fails to go through several measured points not used in the fitting. In particular, it fails to go through the measured optical flux density in several hotspots, which shows the limitation of the 3-fit fitting. Subject to these limitations, however, the SSC model suggests that these hotspots are also, in general, far from equipartition.

Referring to Figure 13, by assuming that the actual infrared flux is that given by the $[3\sigma]$ upper limit for the non-detected hotspots, we find -- as is also the case for the detected sources [see Table 4] -- that most of these hotspots are also far from equipartition.  If the actual infrared flux were, in fact, $\sim$25\% of the adopted upper limit, then this would make little difference from the results shown in Figure 12 and Table 4. $N_{e}$ and $B$ would increase by less than 0.2\%, $B/B_{eq}$ would increase by less than 0.7\%, and $\gamma_{2}$ would decrease by less than 6\%. Thus, the fitting results are quite robust to relatively large changes in the IR flux density.

\subsection{SSC All-Fit Model-Fitting of SCM Hotspots}
To overcome the limitations of the 3-fit model-fitting, for the 8 SCM hotspots with more than 3 measured $SEs$ (with typically a radio $SE$, several IR $SEs$, an optical $SE$ and an X-ray $SE$) we can now also fit for the break in the relativistic electron energy distribution at $\gamma_{b}$ along with the other two parameters ($B$ and $\gamma_{2}$) we determined in the 3-fit model-fitting.  We call this all-fit model fitting. As we explained in Section 4.2 for $\nu_{b} < \nu < \nu_{2}$  and $p=2$ to a good approximation $\alpha_{SE} = 0$ which can reduce the peak in both the synchrotron and IC SEDs. Table 5 shows the results of SSC model fitting using all the measured flux densities for the 8 SCM hotspots. Figure 14 shows the corresponding SED plots. As can be seen the optical data is in general much better fit than with the 3-fit model fitting and the synchrotron peak (and hence $\nu_{2}$) has moved from $\nu_{2} < \nu_{IR}$ to $\nu_{IR} < \nu_{2} < \nu_{O}$ once we fit all the measured SEs and no longer require $\gamma_{b} = \gamma_{2}$. Furthermore, the simplest SO model is completely ruled out for all the 8 SCM hotspots. (See Section 5.1).  As discussed in the next section, although the fit is generally better in the all-fit than in the 3-fit case, the parameters derived in both fits, notably $B$ and $N_{e}$, agree well in the two cases. The reason for this is that both $B$ and $N_e$ are strong functions of $S_R$, $S_X$, $p$, and $\gamma_1$, but  only weak functions of $S_{IR}$, $\gamma_b$ and $\gamma_2$ and both cases use the same values of $S_R$, $S_X$, $p$, and $\gamma_1$ in the  model-fitting.

\subsection{SSC: Comparison of 3-Fit and All-Fit Model-Fitting}
For the SCM hotspots we can compare the results of our 3-fit model-fitting (in which we determined $B$ and $\gamma_{2}$ and derived $N_{e}$) with the all-fit model-fitting (in which we also determined $\gamma_{b}$) for the same sources. This will give us an estimate of what confidence we can place in the 3-fit model-fitting for the SWM hotspots where there are only 3 flux densities to fit. Figure 15 shows how the 3-fit and all-fit SCM hotspot model fitting results compare. The 3-fit results overestimate $B$ by at most 2.5\% and underestimate $N_{e}$ by at most 11\%. Also, the 3-fit $\gamma_{2}$ is lower than the all-fit $\gamma_{2}$ by typically a factor of 3. Thus, the 3-fit $\gamma_{2}$ is seen to be a compromise value when all-fit is not used or can't be used as in the case of the SWM hotspots. For all SCM hotspots, all-fit $\gamma_{b}  < $ 3-fit $\gamma_{2} < $ all-fit $\gamma_{2}$ which is a natural consequence of this compromise. Also, when all-fit is used the higher frequency turnover $\nu_{2}$ is above $\nu_{IR}$ but below $\nu_{O}$ whereas when 3-fit is used $\nu_{2}$ is less than $\nu_{IR}$. What these comparisons show is that both $B$ and $N_{e}$ can be reasonably well determined with the 3-fit modeling even if more infrared flux densities are available. Again, this is explained by the fact that the $B$ and $N_e$ depend primarily on $S_R$, $S_X$, $p$, and $\gamma_1$, which also means that if a hotspot IR flux density is contaminated by lobe emission which will increase its observed $SE_IR$ this will have a small impact on the derived $B$ and $N_e$ parameters. For example if the $SE_IR$ in 3C330 N ($z=0.549$) is decreased by a factor of 8 (assuming that $7/8$ of its IR emission comes from the lobe and not from the hotspot itself which is quite an extreme assumption) then the fitted values of $B$ and $N_e$ donÕt change (to the 2 decimal places we quote) but $\gamma_2$, $EDR$, $B/B_{eq}$ change by factors of only 0.84, 1.03, and 1.01 respectively.  Thus, in general, lobe contamination for the IR flux densities will have little impact on our model-fitted parameters in particular $B$ and $N_e$. By  comparison, if we use $\gamma_1 = 500$ instead of $\gamma_1 = 1000$ in our 3-fit model-fitting for 3C330 N then the fitted values of $B$, $N_e$, $\gamma_2$, $EDR$, $B/B_{eq}$ change by factors of 1.25, 1.42, 0.89, 1.98, and 1.22 respectively compared to their $\gamma_1 = 1000$ values. Under the approximation that $N_e = K_e/\gamma_1$ for $p=2$ reducing $\gamma_1$ by a factor of 2 is equivalent to increasing $S_R$ by a factor of 2 but keeping $\gamma_1$ = 1000. If this approximation were exactly true we would expect that the ratio of the product $N_eB^{3/2}$ for $\gamma_1 = 500$ to $\gamma_1 = 1000$ would be exactly 2. In fact the value of this ratio derived from the model-fitting is 1.98.

\subsection{SSC: Comparison of our 3-fit Results with H04's Results}
An informative way to compare our results with H04 is to plot our predicted value of $R_{H04}$ namely $\left(B/B_{eq}\right)^{-3/2}$ (cf. equation (5)) against H04's $R_{H04}$, as shown in Figure 16. As can be seen there is a good agreement between our 3-fit model fitting using our SSC model (based on the G09 SSC model) and the H04's $R_{H04}$ value derived by H04 using their own SSC model. Of the 24 hotspots we studied we find only one, 3C280W, has a magnetic field above equipartition. Note that those hotspots in which IR limits were used in the model fitting show the same trend as those hotspots for which there are IR detections.

\section{Emission Model Discussion}
\subsection{Is the SO Model Viable?}
For the most straightforward SO model to be viable, for a given hotspot, all the SEs for that hotspot from the radio to X-ray must be consistent with the synchrotron emission theory presented in Section 4.2. That is, $\alpha_{SE}(\nu_{1}<\nu<\nu_{b})$  can vary between $-0.5$ and $0$ (for $2\leq p \leq 3$) and furthermore $\alpha_{SE}(\nu_{b}<\nu<\nu_{2}) = \alpha_{SE}(\nu_{1}<\nu<\nu_{b}) + 0.5$ and so can vary between $0$ and $0.5$ (for $2\leq p \leq 3$).  Also for $\nu>\nu_{2}$, $\alpha_{SE}(\nu>\nu_{2}) \gg 0.5$ since the synchrotron emission decreases very rapidly when $\nu>\nu_{2}$. With only measured $SE_{R}$ and $SE_{X}$ values, and either an upper limit on or a measured value of $SE_{O}$, and with complete freedom to choose $\nu_{b}$ and $\nu_{2}$, and with $p$ between 2 and 3 the SO model is valid for many hotspots. When H04 wrote their paper these conditions were met for many hotspots and hence a SO model was a plausible model at this time, and this is the model they advocated, especially for those hotspots far out of equipartition with $B \ll B_{eq}$. However, adding one or more IR $SEs$ for each hotspot provides extra constraints that the SO model must satisfy to remain valid. In fact, our IR data completely rule out the SO model for all 8 SCM hotspots (as clearly shown in Figures 12 and 14) and also for 6 out of 9 SWM hotspots with measured IR flux densities (see Figure 11). It is not possible to fit a broken power law which satisfies the above constraints to the full dataset for these objects. The 3 SWM hotspots with measured IR flux densities for which a SO model remains valid are 3C109 S, 3C321 E and 3C452 W. Even for the seven hotspots with only IR upper limits a SO model cannot be ruled out for only 2 of our hotspots (3C321 W and 3C324 W). See Figure 13 for their SEDs.

Thus, whereas H04 concluded that the SO model was valid for all hotspots with the exception of a few hotspots near equipartition, we now find that it can be ruled out for all but 5 of the 24 hotspots we studied. By contrast, the SSC model can account for the multi-waveband emission from all the hotspots (but at the price of the majority of hotspots having $B$ well below the equipartition values). Under these circumstances, it seems appropriate to discuss several other possible models.

\subsection{Are SEC Models Viable?}
\subsubsection{SEC-CMB models}
SEC models are ones in which the seed photons for upconversion by Compton scattering come from an external source. The most common SEC model is one in which the CMB is the source of the external photons and we call this model SEC-CMB. SEC-CMB has become the standard explanation for X-ray emission in relativistic extragalactic jet components that point close to the line of sight and thus give the CMB photons a very large Doppler boost (see Worrall (2009) for an excellent review on this subject). However, for hotspots in FR II radio galaxies and lobe-dominated quasars for which there is no suggestion of jets it seems very unlikely that SEC-CMB is a significant component to the X-ray flux density for two reasons:

1.	Large angles ($\theta_{H} \approx 60\deg$) of the hotspots' motion to the line of sight.

2.	Low hotspot advancement speeds: $\beta_{H} \approx 0.1$ (although material in the hotspots can be moving faster than this).

Under these circumstances the Doppler boost is negligible and SEC-CMB is unable to explain hotspot X-ray emission.

\subsubsection{SEC-UC}
An alternative SEC model proposed by Georganopoulos and Kazanas (2003) [G\&K(03)] is that after passing thru the terminal shock the shocked jet material decelerates, and that upstream synchrotron emitted photons from slower moving post terminal shock material can then get Inverse Compton scatted by the relativistic electrons located near the shock front. G\&K(03)  call this mechanism Upstream Compton (UC)  scattering and it is a form of SEC. Hence, we call it SEC-UC. While this mechanism has some attractive features which could allow the hotspot region near the terminal shock to be in equipartition, we are unable to test it with the present dataset. With improved astrometry, X-ray- and radio-emitting regions would be found to be offset from one another with the X-ray emitting regions closer to the shock front if this model applies.

\subsubsection{Implications of Pic A Observations and the SO Compact-Diffuse Hotspot Model}
The different radio and X-ray spectral indices of Pic A W mentioned in Section 4.2 show that the SSC model for this source in its simplest form cannot be correct. Furthermore, the Tingay et al.\ (2008) VLBI observations of the Pic A W hotspot (hereinafter, T08) show this hotspot contains 5 compact radio regions and that the observed X-ray emission could be produced primarily from these regions as synchrotron radiation. We call this the SO Compact-Diffuse (SO-CD) hotspot model   In the SO-CD model a diffuse hotspot is responsible for producing the majority of the radio, IR, and optical emission and a minority of the X-ray emission (via IC emission). Embedded in this diffuse hotspot is a compact hotspot region responsible for producing the minority of the radio, IR, and optical emission but the vast majority of the X-ray emission via synchrotron radiation. This model further assumes that both the diffuse and compact hotspot regions are in equipartition with the same magnetic field strength (350 $\mu G$ in the case of Pic A W).  This assumption is justified since the equipartition B-field is proportional to $(S_{R}[Jy]/\theta_{H}^{3}[mas])^{2/7}$ for $p=2$. For example, for the diffuse hotspot this ratio is $3.0\times 10^{-3}$ (at 6 cm for the radio flux density) while for the 5 compact hotspots listed in Table 1 of T08 this ratio (also at radio wavelength of 6 cm) varies between $3.6\times 10^{-3}$ and $4.5\times 10^{-3}$ taking into account the different observing wavelengths for the compact VLBI hotspots (18 cm) compared to the H04 radio data (6 cm) and assuming $\alpha_{\nu} = 0.5$. One further assumption in this model is that the break frequency $\nu_{b}$ for the compact hotspots is in the X-ray regime due to the fact that the effective radiative lifetime ($\tau$) associated with the young dynamic compact hotspot regions is the same as their dynamical timescale $\sim10$ years compared to $\sim500$ years for the radiative lifetime in the diffuse hotspot regions (See T08 for more details). Since the break frequency scales as $\tau^{-2}$ this allows $\nu_{b}$ to be $\sim 2,500$ times higher for the compact hotspot regions compared to the diffuse hotspot and consequently these regions can then produce sufficient  synchrotron X-ray emission to account for all the observed X-ray flux density.

In Appendix A we examine the SO-CD model further and show that it could account for the hotspots in our sample that appear to be furthest from equipartition, but not for those which appear closest to equipartition.  Thus, at the two extremes ($B/B_{eq} \ll 1$ and $B/B_{eq} \approx 1$) two different emission models are required: namely SO-CD (at least in the case of Pic A W) and SSC. Unfortunately, unlike Pic A W, for the other hotspots in our sample we do not have VLBI data or X-ray spectral data that could easily distinguish between the SO-CD and SSC models. Consequently both remain possible contenders for the majority of hotspots in our sample. 

\subsubsection{Emission Model Conclusions}
The SSC, SO-CD, and SEC-UC models appear to provide possible explanations of the data on our hotspots. At the moment, while SEC-UC has some attractive features to it, there is little in the way of current data to constrain it. So far there is only one hotspot for which we have data to prefer the SO-CD model over the SSC model, namely Pic A W.  Consequently, in the next few subsections we examine the consequences of assuming that the SSC model applies in all other cases.

\subsection{How Can Hotspots Be So Far Out of Equipartition?}
A very interesting result for the hotspots we studied, if the SSC model is correct, is that we find only one hotspot has $B/B_{eq} > 1$ ( = 1.35 for 3C 280 W, or just above equipartition), and all the rest have $B/B_{eq} < 1$, with 10 out of 17 with infrared detections having $B/B_{eq} < 0.1$. Thus, hotspots have $B/B_{eq}$ well below their equipartition value and furthermore something is preventing them from having $B/B_{eq} > 1$. A very natural explanation for this result was proposed by DeYoung (2002). In a study of magnetic field amplification in FR II hotspots he used 3D MHD hotspot magnetic field simulations to conclude that magnetic field amplification by turbulence to equipartition values  (i.e. $B/B_{eq} = 1$) is reached only for a subset of possible initial conditions and that the time for amplification to equipartition may well exceed the dwell time of the fluid in the hotspots unless special conditions are imposed. In other words, $B/B_{eq} < 1$ for most hotspots and only under special conditions does $B/B_{eq}$ approach unity. These simulations are very consistent with our observational results and therefore it seems to us that the magnetic field amplification model by turbulence as developed by De Young (2002) is a very plausible physical model to explain our results when the SSC model applies. 

\subsection{Correlations Among Hotspot Physical Parameters}
Figure 17 shows how various physical parameters that are related to the magnetic field and relativistic electron energy density (determined from the 3 flux density fitting) are related to one another and to the ratio $B/B_{eq}$, which shows how far the magnetic field is out of its equipartition value.  The top right plot shows that there is relatively small spread in $B_{eq}$. This roughly constant value of $B_{eq}$ is a consequence of synchrotron theory in which $B_{eq}$ is a weak function of both the hotspot radio luminosity ($L_{R}$) and volume ($V_{h}$). For example, for $p=2$, $B_{eq}$ depends on the ratio $\left(L_{R}/V_{h}\right)^{2/7}$. This automatically leads to the strong correlation between $B$ and $B/B_{eq}$ shown in the top left plot. The strong correlation in the bottom right plot can be simply explained by the equipartition condition in which, for $p=2$ and $\gamma_{1}$ = constant (= 1000 in our modeling), $N_{eq}$  and $B_{eq}$ are related by $N_{eq} = \eta(\gamma_{2})B_{eq}^{2}$ where $\eta(\gamma_{2}) = (1-\gamma_{1}/\gamma_{2})/(2\mu_{0}\gamma_{1}m_{e}c^{2}\ln(\gamma_{2}/\gamma_{1}))$. For the $\gamma_{2}$ values listed in Table  4, $\eta(\gamma_{2})$ varies by only a factor of 1.51, which leads to the very strong $B_{eq}-N_{eq}$ correlation shown in this plot. Combining $N_{eq} = \eta(\gamma_{2}) B_{eq}^{2}$ with equation (5) leads to $B = \eta(\gamma_{2})B_{eq}^{7/2}N_e^{-3/2}$, and this accounts for the anti-correlation between $B$ and $N_{e}$ shown in the bottom left plot of Figure 18. Equation (3) shows that $B/B_{eq}$ is also proportional to $(\nu_{2}/B_{eq})\gamma_{2}^{-2}$, and the model-fitting results also show that $\nu_{2}$ is approximately constant ($\approx 10^{13.5} Hz$), which leads to a strong anti-correlation between $B/B_{eq}$ and $\gamma_{2}$. Thus, correlations between $B/B_{eq}$ and other physical parameters need to be treated with caution, as they are often a natural consequence of synchrotron theory and of the fact that $B_{eq}$ and $\nu_{2}$ are approximately constant. $\nu_{2}$ being in the IR naturally leads to there being negligible X-ray synchrotron radiation.

The plots in Figure 18 show how that both the 5 GHz K-corrected radio luminosity ($L_{R}$) and redshift ($z$) are both strongly correlated with $B/B_{eq}$ and hence also with each other.  The first of these plots was also effectively shown by H04, who showed that $L_{R}$ and $R_{H04}$ are anti-correlated. From equation (5) (since $R_{H04}$ and $B/B_{eq}$ are related by $R_{H04} = \left(B/B_{eq}\right)^{3/2}$), for $p=2$ this must lead to a correlation between $L_{R}$ and $B/B_{eq}$.

\section{Comparison with Other Recent Studies of Hotspots}
Our basic result is that the hotspots we studied are not well fit by a simple single-zone SSC model in which the particle and field energies are close to equipartition but instead require that the magnetic field is often substantially weaker than the equipartition value if the SSC model is applicable.  We note that this result could be due in part to selection effects because we limit ourselves to hotspots with X-ray detections [see also H04]; for a simple SSC model, a hotspot with $B<B_{eq}$ will have X-ray emission considerably brighter than the corresponding equipartition value. Nevertheless, it is interesting to compare our results with those of other recent studies of hotspots. As already noted, the well-studied Cyg A and PKS 1421-490 hotspots -- both detected in X-rays -- are in fact near equipartition (Stawarz et al.\ 2007; Godfrey et al.\ 2009). This is of course in no way inconsistent with our results, which show that most X-ray detected hotspots are far from equipartition.

Kraft et al.\ (2007) have published a multi-wavelength study of the hotspots in the nearby radio galaxy 3C33 $(z=0.06)$ which includes Spitzer data at all four IRAC bands and at MIPS 24 $\mu m$.  They isolate the emission from six regions within the extended lobe/hotspot complexes at the tips of the radio lobes of this galaxy.  The brightest of these has a radio-to-X-ray SED consistent with an equipartition $B$ field.  The other five regions are far from equipartition, with the $B$ fields about an order of magnitude weaker than $B_{eq}$, and the X-ray emission correspondingly stronger, as we report for most of our targets. They suggest that the X-ray emission in these regions is due to synchrotron emission while pointing out that this requires multiple populations of relativistic electrons. This agrees with our conclusion that a single broken power law energy distribution cannot account for the observations. These findings are different from Tingey et al.\ (2008) who found that the most compact region of Pic A W  is well out of equipartition if the SSC model is correct and only radio data on the arcsecond-scale is used to estimate the emission region size. In contrast, for 3C 33 Kraft et al.\ (2007) find that the most compact hotspot component on the arsecond-scale is in equipartition. However, Tingay et al.\ (2008) also found that the most compact region of the Pic A W hotspot contains several VLBI-scale radio components which if they are in equipartition could produce the observed X-ray flux density of this region which gave rise to the SO-CD model we described earlier. Both sets of observations are consistent with compact regions producing SO emission which is responsible for the majority of the X-ray emission in certain hotspot regions. It is just that these location regions relative to the primary hotspot (as defined by Kraft et al.\ (2007)) differ between the Pic A W and 3C 33 hotspots.

Zhang et al.\ (2010) have compiled existing  X-ray, visible, and radio data on a number of hotspots and come to conclusions similar to those that we present.  They show that a simple synchrotron model does not work in most cases and that, in the absence of relativistic bulk motion, an SSC interpretation requires that many hotspots have magnetic fields far below the value required for equipartition.  Only 8 of the 17 hotspots for which we have detections in all three spectral bands are included in their sample.

Finally, Mack et al.\ (2009) present optical, infrared, and radio observations of a sample of nine low-power radio hotspots, which, in general, do not also have X-ray detections, and report detecting at least four of the targets at J, H, and/or K.  Their analysis suggest a high break frequency ($10^{5}$  to $10^{6}$ GHz) in the SEDs of these hotspots, attributable to the relatively low magnetic fields appropriate for these low-power hotspots.  However, in the absence of the X-ray data, their results cannot be directly compared with ours.

\section{Conclusions}
We present \emph{Spitzer} measurements -- primarily in the IRAC bands from 3.6-to-8 $\mu m$ -- of 18 hotspots in the extended lobes of radio sources as well as upper limits to the infrared fluxes of 9 more. We analyze in detail a subset of 17 infrared- and X-ray-detected hotspots and 7 with infrared upper limits for which Hardcastle et al.\ (2004) also report X-ray detections. For a simple one-zone synchrotron self-Compton [SSC] model of the spectral energy distributions, only 1 of the 17 detected hotspots has a magnetic field strength greater than the equipartition value, 1 has a field which is less than 25\% below the equipartition value, while 10 of the 17 have fields which are one-tenth or less of the equipartition value. We point out that the departures from equipartition are qualitatively consistent with a suggestion by DeYoung (2002) that the magnetic fields in hotspots are amplified to the equipartition value by turbulence, and that the dwell time of the field in the hotspot may not be long enough for the field to reach equipartition in this fashion. Hardcastle et al.\ (2004) had previously concluded that the X-ray detected hotspots are, in general, far from equipartition if a simple SSC model is adopted. Our infrared data allow us, in addition, to rule out straightforward synchrotron only (SO) models for the hotspot emission. We caution the reader, however, that because we have observed only hotspots with X-ray detections, our conclusions may not apply to all hotspots, as X-rays should be weak in hotspots that are in equipartition.  

For the hotspot in our sample with the best and most complete data, Pic A W, the discrepancy between the radio and X-ray spectral indices rules out the SSC interpretation. Based on VLBI imaging which resolves compact substructure in the lobes of this nearby source, T08 have proposed a hybrid model for Pic A W in which the radio, IR, and optical emission come predominantly from the background diffuse hotspot while the X-rays are produced by synchrotron emission from the compact regions seen in the VLBI data.  Both regions are in equipartition with the same magnetic field, albeit with different particle populations.   This model, which we have christened SO-CD, may be a viable alternative to the SSC model for those sources that are far from equipartition in that picture.  However, verifying this conjecture requires VLBI imaging, which may not be practical for hotspots that are fainter and more distant than Pic A W.  

We thank the referee and Martin Hardcastle for their thoughtful comments, which improved the paper significantly. This work is based on observations made with the \emph{Spitzer} \emph{Space Telescope}, which is operated by the Jet Propulsion Laboratory, California Institute of Technology under a contract with NASA. Support for this work was provided by NASA through an award issued by JPL/Caltech.   

\appendix
\section{The Synchrotron-Only Compact-Diffuse Model}
We can examine the SO-CD model further and see under what conditions it might apply to hotspots other than Pic A W in our sample. If we assume that in each hotspot there is a compact VLBI-scale hotspot emission region, which has radio and X-ray flux densities of $F_{R}$, $F_{X}$ with angular size of $\theta_{C}$, and that this compact hotspot is in equipartition, then assuming $p=2$:
\begin{equation}
F_{R} = c_{R} B_{eq}^{7/2} \theta_{C}^{3}
\end{equation}
Furthermore, if like Tingay et al.\ (2008) [T08], we assume that this compact hotspot has a break frequency in or above the X-ray regime and that Inverse Compton emission is negligible for all frequencies at or below this break then it also follows that:
\begin{equation}
F_{X} \approx F_{R} \left(\frac{\nu_{R}}{\nu_{X}}\right)^{1/2}
\end{equation}
($\approx S_{X}$ if this component dominates the X-ray emission), where $c_{R}$ is constant for a given hotspot.  Also, like T08, if we also adopt the assumption that the diffuse and compact hotspots are in equipartition with the same magnetic field strengths then the compact hotspot radio flux density, $F_{R}$,  can be related to the diffuse hotspot flux density, $S_{R}$, by the ratio $\beta = F_{R}/S_{R}$ as follows:
\begin{equation}
F_{R} = \beta S_{R} = \beta c_{R} B_{eq}^{7/2} \theta_{D}^{3}
\end{equation}
under the assumption that $c_{R}$  is the same for both the compact VLBI-scale hotspot region and the more diffuse hotspot region (which emits most of the radio, IR, and optical flux density) and has an angular size of $\theta_{D}$.  Combining the above equations we get:
\begin{equation}
\beta \approx \left(\frac{\nu_{R}}{\nu_{X}}\right)^{1/2} \left(\frac{SE_{X}}{SE_{R}}\right)
\frac{\theta_{C}}{\theta_{D}} = \beta^{1/3} \approx \left(\frac{\nu_{R}}{\nu_{X}}\right)^{1/6} \left(\frac{SE_{X}}{SE_{R}}\right)^{1/3} 
\end{equation}
For the radio and X-ray frequencies we use, $(\nu_{R}/\nu_{X})^{1/6} \approx 1/20$. Hence:
\begin{equation}
\frac{\theta_{C}}{\theta_{D}} \approx \frac{1}{20} \left(\frac{SE_{X}}{SE_{R}}\right)^{1/3} 
\end{equation}
In Figure 19 the CD-SO model parameters $\beta$ and $\theta_{C}$ are plotted against the SSC model parameter $B/B_{eq}$. In both plots the western hotspot of Pic A (which lies in the closest FR II source) occupies the top left of each plot. As can be seen both $\beta$ and $\theta_{C}$ are anti-correlated with $B/B_{eq}$, which isn't surprising as it says that those hotspots which have the most X-ray emission above the equipartition value require compact components that produce a larger fraction of the total radio flux density and consequently have larger angular sizes as well. However, the CD-SO model cannot explain those hotspots close to equipartition ($B/B_{eq} \approx 1$) since in those cases the diffuse hotspot is close to equipartition and IC emission from the diffuse region can explain all the observed X-ray emission without the need for any additional CD-SO emission. Thus, the CD-SO model has its limitations as well.

\begin{deluxetable}{lccccccc}
\tabcolsep 3.8pt
\tablecaption{Hotspot Observations}
\tablehead{
\colhead{Name} 	&	Type	&	Hotspot RA	&	Hotspot Dec	&	Redshift		&	Angular Size	&	R	&	Comment	\\
&	&	(J2000)	&	(J2000)	&	(z)	&	(arcsec)	&	&	}
\startdata
3C 6.1	N	&	SWM	&	00h16m33.3s	&	+79d17m02.3s	&	0.8404	&	0.36	&	2.3	&		\\
3C 6.1	S	&	SWM	&	00h16m29.2s	&	+79d16m39.0s	&	0.8404	&	0.41	&	1.1	&	Confused	\\
3C 47	N	&	SWM	&	01h36m25.8s	&	+20d57m51.8s	&	0.425	&	1.89	&	$<$6.6	&	Bonus	\\
3C 47	S	&	SWM	&	01h36m22.9s	&	+20d56m56.0s	&	0.425	&	0.434	&	14	&	Confused	\\
3C 109	S	&	SWM	&	04h13m42.6s	&	+11d11m38.7s	&	0.3056	&	0.377	&	4.6	&		\\
3C 123	E	&	SWM	&	04h37m04.9s	&	+29d40m10.1s	&	0.2177	&	1.1$\times$0.54	&	2.1	&	Confused	\\
3C 123	W	&	SWM	&	04h37m03.8s	&	+29d40m17.0s	&	0.2177	&	1.0$\times$0.13	&	3.1	&	Confused	\\
Pic A	W	&	SCM	&	05h19m26.2s	&	$-$45d45m54.5s	&	0.03498	&	0.75	&	454	&		\\
3C 173.1	N	&	SWM	&	07h09m19.6s	&	+74d49m58.2s	&	0.292	&	0.26	&	$<$237	&	Bonus	\\
3C 173.1	S	&	SWM	&	07h09m15.3s	&	+74d49m02.2s	&	0.292	&	0.83	&	91	&		\\
3C 179	W	&	SCM	&	07h28m09.8s	&	+67d48m47.5s	&	0.846	&	0.145	&	63	&		\\
3C 207	E	&	SWM	&	08h40m48.0s	&	+13d12m23.2s	&	0.684	&	0.27	&	73	&	Confused	\\
3C 228	N	&	SWM	&	09h50m11.1s	&	+14d20m25.6s	&	0.5524	&	0.203	&	24	&	Confused	\\
3C 228	S	&	SWM	&	09h50m10.6s	&	+14d19m40.0s	&	0.5524	&	0.265	&	31	&		\\
3C 254	W	&	SWM	&	11h14m37.7s	&	+40d37m23.1s	&	0.734	&	0.29	&	8.8	&		\\
3C 263	E	&	SWM	&	11h39m59.4s	&	+65d47m44.0s	&	0.652	&	0.39	&	4	&	Confused	\\
3C 263	W	&	SWM	&	11h39m52.9s	&	+65d47m59.5s	&	0.652	&	0.18	&	$<$11	&	Bonus	\\
3C 265	E	&	SWM	&	11h45m31.2s	&	+31d33m36.2s	&	0.8108	&	0.356	&	2.2	&		\\
3C 265	W	&	SWM	&	11h45m26.4s	&	+31d33m54.8s	&	0.8108	&	0.73	&	15	&	Confused	\\
3C 275.1	N	&	SCM	&	12h43m57.4s	&	+16d23m02.3s	&	0.557	&	1.4$\times$0.2	&	19	&		\\
3C 280	E	&	SWM	&	12h56m57.0s	&	+47d20m19.5s	&	0.996	&	0.186	&	6.7	&		\\
3C 280	W	&	SWM	&	12h56m57.1s	&	+47d20m19.7s	&	0.996	&	0.146	&	0.48	&		\\
3C 295	N	&	SWM	&	14h11m20.4s	&	+52d12m11.9s	&	0.4614	&	0.1	&	1.8	&	Confused	\\
3C 295	S	&	SWM	&	14h11m20.7s	&	+52d12m07.1s	&	0.4614	&	0.1	&	1.1	&	Confused	\\
3C 303	W	&	SCM	&	14h43m00.9s	&	+52d01m39.8s	&	0.141	&	1.1$\times$0.28	&	154	&		\\
3C 321	E	&	SWM	&	15h31m50.5s	&	+24d02m42.0s	&	0.096	&	0.69	&	48	&		\\
3C 321	W	&	SWM	&	15h31m35.8s	&	+24d06m01.0s	&	0.096	&	2.7$\times$0.45	&	210	&		\\
3C 324	E	&	SWM	&	15h49m49.2s	&	+21d25m39.7s	&	1.2063	&	0.365	&	0.93	&		\\
3C 330	N	&	SWM	&	16h09m39.3s	&	+65d56m52.9s	&	0.549	&	0.45	&	0.81	&		\\
3C 330	S	&	SWM	&	16h09m30.3s	&	+65d56m22.8s	&	0.549	&	0.2	&	2.4	&		\\
3C 334	S	&	SWM	&	16h20m22.7s	&	+17d36m11.6s	&	0.555	&	1.34$\times$0.3	&	292	&		\\
3C 351	J	&	SCM	&	17h04m43.8s	&	+60d44m48.2s	&	0.371	&	0.16	&	85	&		\\
3C 351	L	&	SCM	&	17h04m43.5s	&	+60d44m52.7s	&	0.371	&	0.8	&	39	&		\\
3C 390.3	N	&	SCM	&	18h41m47.8s	&	+79d47m44.2s	&	0.0569	&	1.3$\times$0.5	&	1380	&		\\
3C 403	F1	&	SCM	&	19h52m19.1s	&	+02d30m33.0s	&	0.059	&	0.275	&	2149	&		\\
3C 403	F6	&	SCM	&	19h52m17.5s	&	+02d30m33.0s	&	0.059	&	0.256	&	1414	&		\\
3C 452	W	&	SWM	&	22h45m37.8s	&	+39d40m59.0s	&	0.0811	&	0.705	&	356	&
\enddata
\tablecomments{``SWM" are \emph{Spitzer} Warm Mission observations, and ``SCM" are \emph{Spitzer} Cryogenic Mission observations. ``Confused" hotspots are those whose flux could not be adequately distinguished from the flux of the host galaxy in the IRAC images, or whose radio position could not be sufficiently identified with flux in the IR and X-ray images, and are not discussed further in the text. ``Bonus" hotspots are those which were not originally proposed for because they have only an X-ray upper limit, but which we were able to observe.}
\end{deluxetable}

\begin{deluxetable}{lccccccccccccccccc}																		
\tablecaption{Newly Observed with \emph{Spitzer} IRAC}	
\tablehead{																		
\colhead{Name} 	&	1 keV 	&	Optical $\lambda$	&	Optical Flux	&	3.6 $\mu$m	&	5 GHz \\ [-1.5ex]
				&	[nJy]		&	[$\mu$m]		&	[$\mu$Jy]		&	[$\mu$Jy]		&	[mJy]	}
\startdata																		
3C	6.1	N	&	0.45	&	-	&	-	&	$<$1.1		&	340	\\[-1.5ex]				
3C	47	N	&	$<$0.1	&	0.55	&	$<$1.0	&	$<$1.4		&	127	\\[-1.5ex]			
3C	109	S	&	0.15	&	0.67	&	$<$1.4	&	5.1$\pm$0.4		&	181	\\[-1.5ex]				
3C	173.1	N	&	$<$0.12	&	-	&	-	&	5.1$\pm$0.3		&	9	\\[-1.5ex]				
3C	173.1	S	&	0.2	&	0.7	&	$<$1.4	&	$<$1.2		&	33	\\[-1.5ex]				
3C	228	S	&	1.3	&	0.86	&	$<$1.14	&	12.2$\pm$0.4		&	132	\\[-1.5ex]				
3C	254	W	&	0.54	&	0.55	&	$<$0.41	&	33.6$\pm$0.5		&	146	\\[-1.5ex]				
3C	263	W	&	$<$0.06	&	-	&	-	&	$<$1.2		&	23	\\[-1.5ex]				
3C	265	E	&	0.35	&	0.55	&	$<$0.4	&	68.3$\pm$0.2		&	272	\\[-1.5ex]				
3C	280	E	&	0.31	&	0.61	&	0.34	&	$<$6.7		&	82	\\[-1.5ex]				
3C	280	W	&	0.6	&	0.61	&	$<$0.4	&	5.1$\pm$0.2		&	631	\\[-1.5ex]
3C	321	E	&	0.3	&	-	&	-	&	66.3$\pm$0.6		&	125	\\[-1.5ex]
3C	321	W	&	0.12	&	-	&	-	&	$<$1.8		&	20	\\[-1.5ex]
3C	324	E	&	0.2	&	-	&	-	&	$<$2.1		&	277	\\[-1.5ex]
3C	330	N	&	0.35	&	0.55	&	$<$0.6	&	19.4$\pm$0.2		&	625	\\[-1.5ex]
3C	330	S	&	0.068	&	0.55	&	$<$0.6	&	$<$0.6		&	102	\\[-1.5ex]
3C	334	S	&	0.54	&	0.55	&	$<$0.86	&	18.4$\pm$0.3		&	18	\\[-1.5ex]
3C	452	W	&	0.34	&	-	&	-	&	12.2$\pm$0.4		&	33	
\enddata														
\tablecomments{IRAC upper limits are 3-sigma.}														
\end{deluxetable}

\begin{deluxetable}{lccccccccccccccccc}
\centering
\tabcolsep 2.8pt
\tablecaption{Archival \emph{Spitzer} IRAC Data}
\rotate
\tablehead{
\colhead{Name} 	&	1 keV	&	Optical $\lambda$	&	Optical Flux	&	3.6 $\mu$m	&	4.5 $\mu$m	&	5.8 $\mu$m	&	8 $\mu$m	&	5 GHz	\\											
				&	[nJy]		&	[$\mu$m]				&	[$\mu$Jy]		&	[$\mu$Jy]		&	[$\mu$Jy]		&	[$\mu$Jy]		&	[$\mu$Jy]	&	[mJy]	}
\startdata
3C	179	W	&	1.54	&	0.55	&	$<$0.57	&	$<$2.1	&	$<$3.3	&	$<$21.9	&	$<$25.5	&	63	\\
3C	275.1	N	&	1.78	&	0.64	&	0.48	&	-	&	29.3$\pm$3.1	&	-	&	109.5$\pm$21.2	&	191	\\	
3C	303	W	&	4	&	0.55	&	7.5	&	138.7$\pm$0.7	&	124.0$\pm$1.0	&	198.4$\pm$4.6	&	319.0$\pm$4.6	&	257	\\
3C	351	J	&	4.3	&	0.7	&	2.6	&	32.6$\pm$0.5	&	45.5$\pm$0.8	&	96.9$\pm$3.4	&	111.7$\pm$3.6	&	167	\\
3C	351	L	&	3.4	&	0.7	&	2.1	&	24.5$\pm$0.5	&	37.2$\pm$0.7	&	78.4$\pm$3.4	&	106.3$\pm$3.6	&	406	\\	
3C	390.3	N	&	4.5	&	0.67	&	2.6	&	77.5$\pm$0.7	&	82.7$\pm$1.0	&	147.6$\pm$4.6	&	164.8$\pm$4.7	&	87	\\
3C	403	F1	&	1	&	0.7	&	1.08	&	20.4$\pm$1.1	&	28.9$\pm$1.0	&	46.1$\pm$6.3	&	63.8$\pm$5.4	&	21	\\
3C	403	F6	&	1.8	&	0.7	&	2.16	&	44.9$\pm$1.1	&	49.6$\pm$1.1	&	64.6$\pm$3.5	&	85.1$\pm$5.1	&	35	\\	
Pic	A	W	&	89	&	0.67	&	130	&	1305.1$\pm$7.0	&	1657.2$\pm$8.2	&	2209.7$\pm$30.2	&	3264.8$\pm$18.2	&	1930
\enddata																		
\centering																		
\tablecomments{Results of analysis of Archival Spitzer data on hot spots.  Note that 3C 351L was mis-identified as 3C 351K by H04.  We included  the archival Pic A W 24 $\mu m$ flux of 11450 $\mu Jy$ in our analysis.}
\end{deluxetable}

\begin{deluxetable}{lccccc}
\centering
\tablecaption{3-Fit Model-Fitting Results}
\tablehead{
\colhead{Name} 	&	B		&	$\gamma_{2}$	&	$N_{e}$		&	$EDR$	&	$B/B_{eq}$	\\
				&	[$\mu G$]		&					&	[$m^{-3}$]		&				&					}
\startdata
\hline
\multicolumn{6}{|c|}{SWM Hotspots with IR Detections} \\
\hline 															
3C109 S   	&	4.08E+01	&	3.67E+05	&	6.06E-01	&	2.26E-02	&	3.38E-01	\\[0.25 ex]
3C228 S   	&	1.67E+01	&	7.13E+05	&	6.21E+00	&	3.32E-04	&	1.01E-01	\\[0.25 ex]
3C254 W   	&	3.71E+01	&	5.75E+05	&	2.04E+00	&	5.15E-03	&	2.22E-01	\\[0.25 ex]
3C265 E   	&	8.85E+01	&	3.86E+05	&	6.28E-01	&	1.01E-01	&	5.20E-01	\\[0.25 ex]
3C280 W   	&	6.32E+02	&	1.03E+05	&	1.47E+00	&	2.83E+00	&	1.35E+00	\\[0.25 ex]
3C321 E   	&	5.44E+00	&	1.39E+06	&	1.94E+00	&	1.02E-04	&	7.25E-02	\\[0.25 ex]
3C330 N   	&	1.29E+02	&	2.27E+05	&	2.79E-01	&	5.31E-01	&	8.34E-01	\\[0.25 ex]
3C334 S   	&	1.77E+00	&	3.41E+06	&	1.14E+01	&	1.64E-06	&	2.22E-02	\\[0.25 ex]
3C452 W   	&	9.01E-01	&	3.17E+06	&	7.90E+00	&	6.20E-07	&	1.68E-02	\\[0.25 ex]

\hline 															
\multicolumn{6}{|c|}{SCM Hotspots with IR Detections} \\
\hline 															
3C275.1 N 	&	1.56E+01	&	6.95E+05	&	4.43E+00	&	4.07E-04	&	1.07E-01	\\[0.25 ex]
3C303 W   	&	5.72E+00	&	1.39E+06	&	1.55E+01	&	1.42E-05	&	4.12E-02	\\[0.25 ex]
3C351 J   	&	1.51E+01	&	7.82E+05	&	3.39E+01	&	4.88E-05	&	5.86E-02	\\[0.25 ex]
3C351 L   	&	7.15E+00	&	9.73E+05	&	2.02E+00	&	1.78E-04	&	8.49E-02	\\[0.25 ex]
3C390.3 N 	&	6.70E-01	&	4.41E+06	&	6.04E+01	&	4.30E-08	&	7.86E-03	\\[0.25 ex]
3C403 F1  	&	8.08E-01	&	4.11E+06	&	1.24E+02	&	3.07E-08	&	7.13E-03	\\[0.25 ex]
3C403 F6  	&	1.16E+00	&	3.73E+06	&	1.49E+02	&	5.33E-08	&	8.36E-03	\\[0.25 ex]
PicA W    	&	3.77E+00	&	1.72E+06	&	8.48E+01	&	1.09E-06	&	1.98E-02	\\[0.25 ex]

\hline 															
\multicolumn{6}{|c|}{Hotspots with IR upper limits} \\
\hline 															
3C6.1 N   	&	1.02E+02	&	2.31E+05	&	6.44E-01	&	1.42E-01	&	5.73E-01	\\[0.25 ex]
3C173.1 S 	&	1.53E+00	&	1.93E+06	&	1.42E+00	&	1.06E-05	&	3.79E-02	\\[0.25 ex]
3C179 W   	&	1.87E+01	&	6.55E+05	&	2.34E+01	&	1.12E-04	&	7.44E-02	\\[0.25 ex]
3C280 E   	&	6.05E+01	&	4.18E+05	&	3.10E+00	&	9.48E-03	&	2.64E-01	\\[0.25 ex]
3C321 W   	&	9.10E-01	&	2.56E+06	&	8.19E+00	&	6.25E-07	&	1.69E-02	\\[0.25 ex]
3C324 E   	&	1.90E+02	&	1.97E+05	&	3.40E-01	&	9.68E-01	&	9.91E-01	\\[0.25 ex]
3C330 S   	&	1.03E+02	&	2.19E+05	&	7.22E-01	&	1.33E-01	&	5.61E-01
\enddata
\centering
\tablecomments{3-Fit Model-Fitting Results. 
Fitted parameters $B$ and $\gamma_{2}$ as well as the derived parameters $N_{e}$, $EDR$, and $B_{eq}$ obtained by fitting $SE_{R}$ and $SE_{X}$ and applying the constraint $S_{R} = c_{R} K_{e} B^{3/2}$ for the 24 hotspots in our sample. Note that when fitting the hotspots with IR 3.6 $\mu m$ upper limits the upper limit value of was used  for $S_{IR}$ to determine the value quoted and 4 times lower than the upper limit was used in the fitting to determine if this value was an upper or lower limit.}
\end{deluxetable}

\begin{deluxetable}{lccccccccccccccc}
\centering
\tablecaption{SCM Hotspots -- All-Fit Model-Fitting Results}
\tablehead{
\colhead{Name} 	&	B				&	$\gamma_{b}$	&	$\gamma_{2}$	&	$N_{e}$		&	$EDR$	&	$B/B_{eq}$	\\
					&	[$\mu G$]		&						&						&	[$m^{-3}$]		&				&					}
\startdata
3C275.1 N 	&	1.52E+01	&	4.70E+04	&	1.99E+06	&	4.77E+00	&	4.84E-04	&	1.13E-01	\\
3C303 W   	&	5.69E+00	&	1.92E+05	&	4.19E+06	&	1.57E+01	&	1.61E-05	&	4.27E-02	\\
3C351 J   	&	1.48E+01	&	6.89E+04	&	2.42E+06	&	3.53E+01	&	5.79E-05	&	6.16E-02	\\
3C351 L   	&	6.99E+00	&	3.30E+04	&	3.02E+06	&	2.37E+00	&	2.20E-04	&	9.02E-02	\\
3C390.3 N 	&	6.65E-01	&	1.50E+06	&	5.26E+06	&	6.05E+01	&	4.42E-08	&	7.92E-03	\\
3C403 F1  	&	8.07E-01	&	1.50E+06	&	1.24E+07	&	1.25E+02	&	3.10E-08	&	7.16E-03	\\
3C403 F6  	&	1.16E+00	&	1.23E+06	&	1.11E+07	&	1.50E+02	&	5.44E-08	&	8.40E-03	\\
PicA W    	&	3.75E+00	&	3.01E+05	&	5.62E+06	&	8.54E+01	&	1.20E-06	&	2.04E-02	\enddata
\centering
\end{deluxetable}

\clearpage
\begin{figure}
\includegraphics[width=0.99\textwidth]{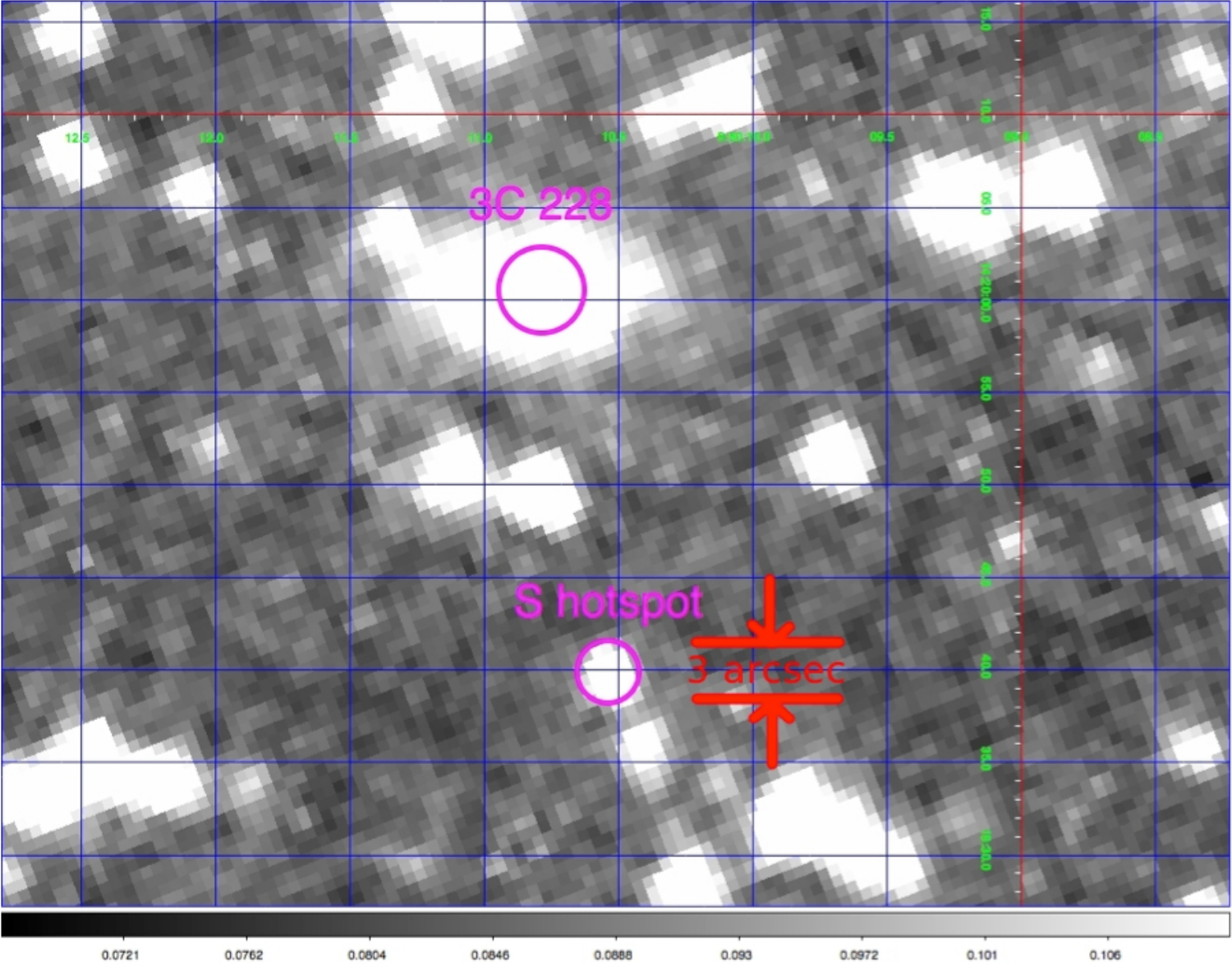}
\caption{3C 228 S hotspot. \emph{Spitzer} 3.6 $\mu m$ Image. In these images, the smaller purple circle is at the position of the radio hotspot, whereas the larger purple circle is centered on the position of the core radio emission.}
\end{figure}

\clearpage
\begin{figure}
\includegraphics[width=0.99\textwidth]{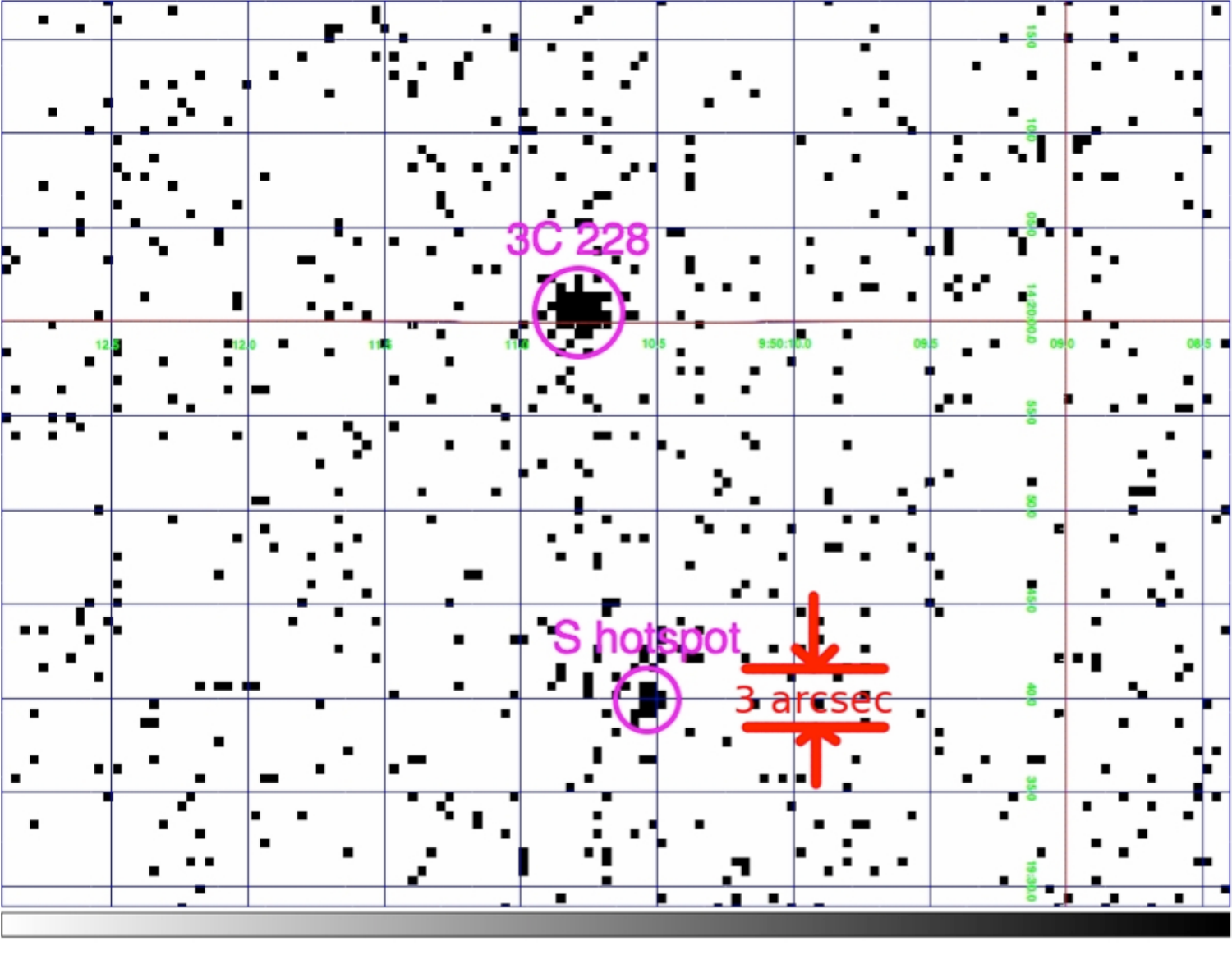}
\caption{3C 228 S hotspot. \emph{\emph{Chandra}} Image.}
\end{figure}

\clearpage
\begin{figure}
\includegraphics[width=0.99\textwidth]{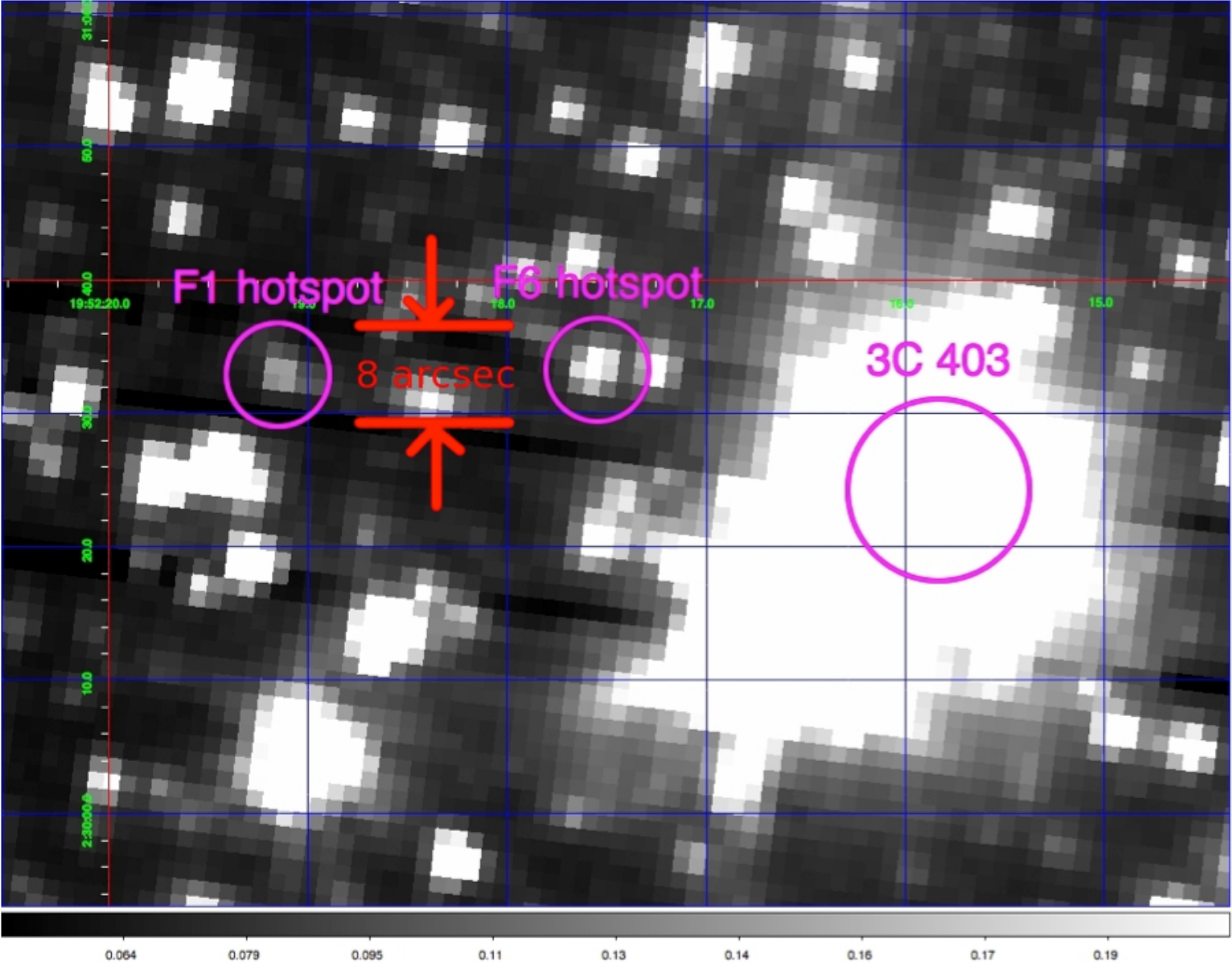}
\caption{3C 403 F1 and F6 hotspots, \emph{Spitzer} 3.6 $\mu m$ image.}
\end{figure}

\clearpage
\begin{figure}
\includegraphics[width=0.99\textwidth]{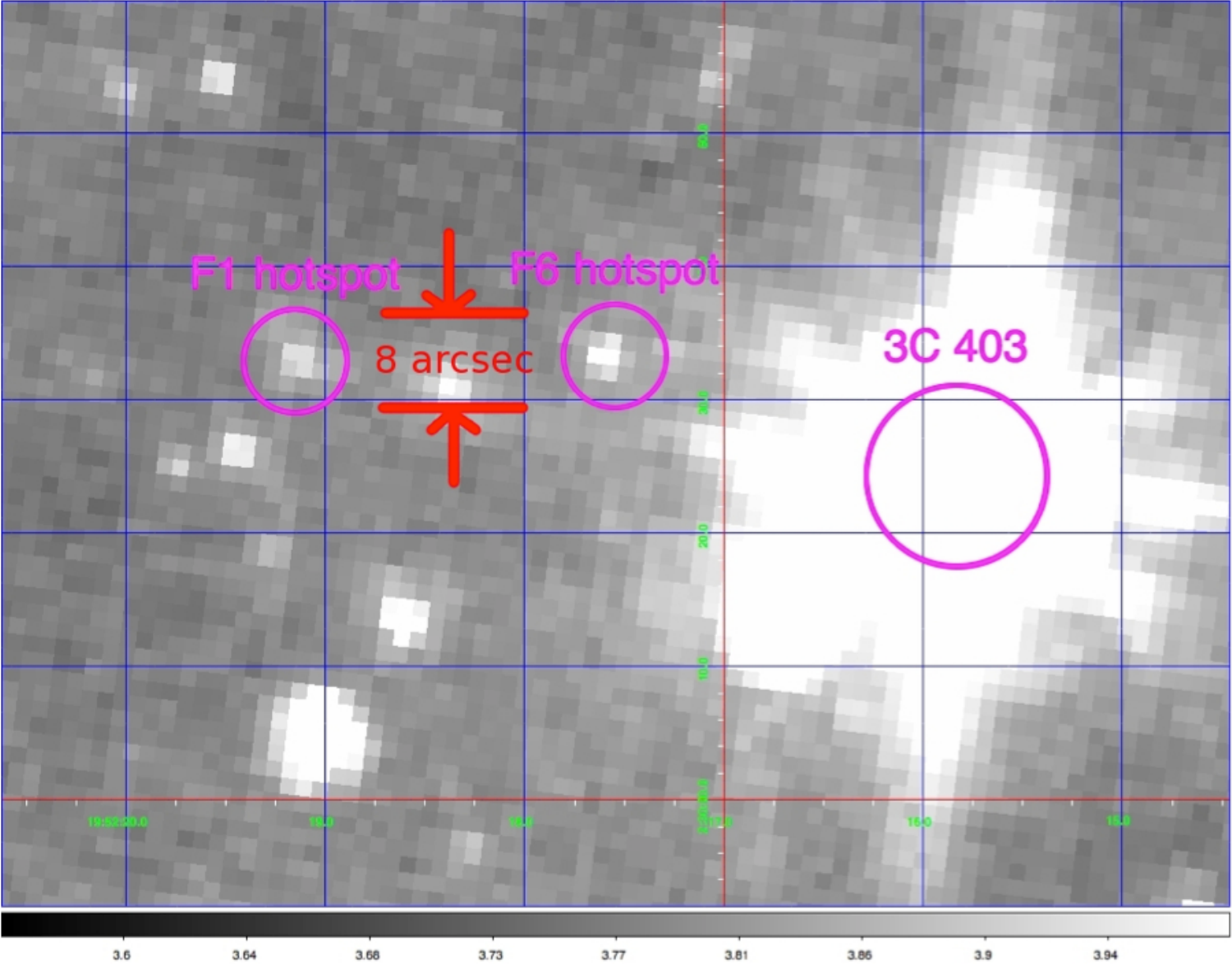}
\caption{3C 403 F1 and F6 hotspots, \emph{Spitzer} 8 $\mu m$ image. Note that at 8 $\mu m$ the hotspots stand out relative to the stars in the image because of their red SEDs.}
\end{figure}

\clearpage
\begin{figure}
\includegraphics[width=0.99\textwidth]{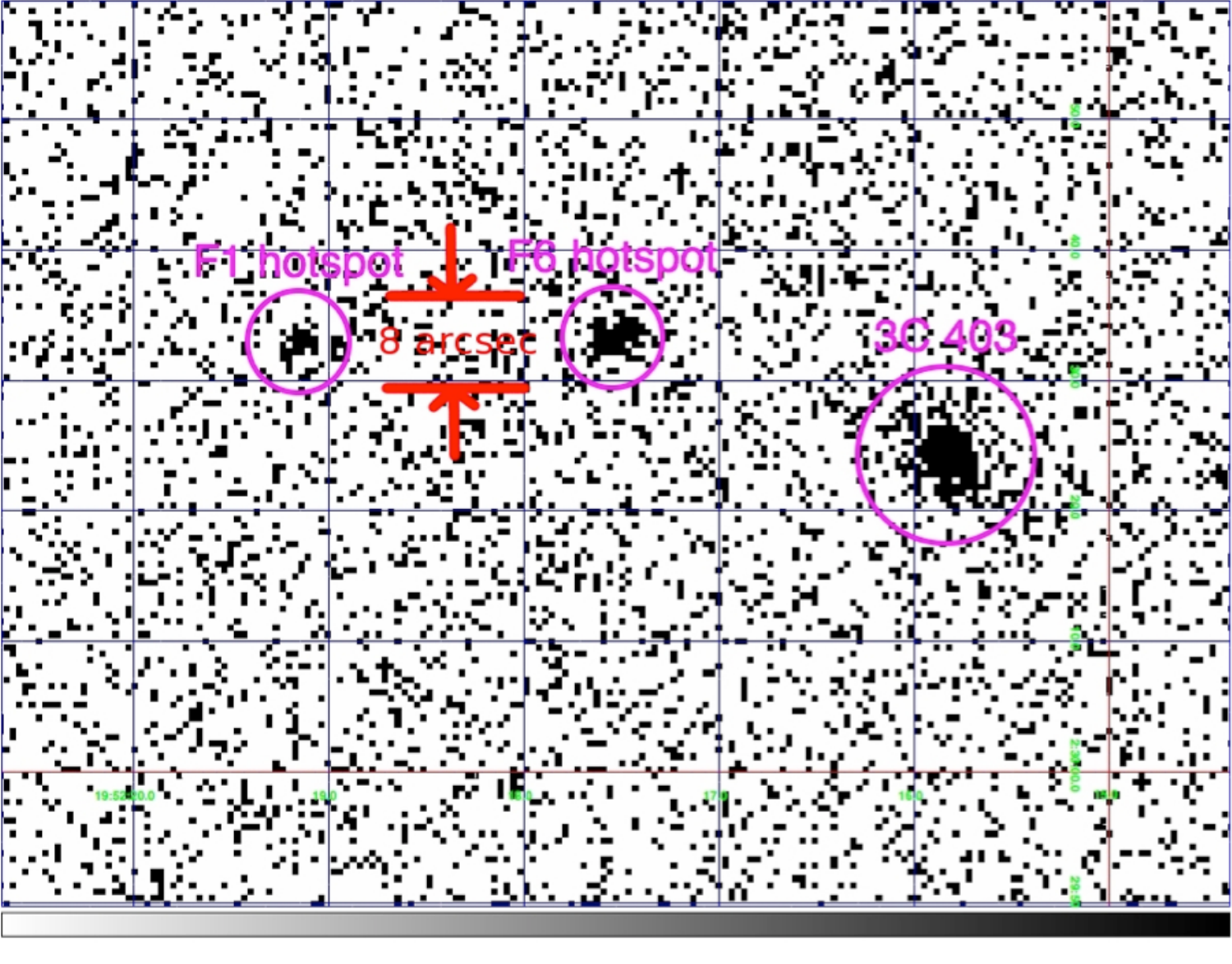}
\caption{3C 403 F1 and F6 hotspots, \emph{Chandra} image.}
\end{figure}

\clearpage
\begin{figure}
\includegraphics[width=0.99\textwidth]{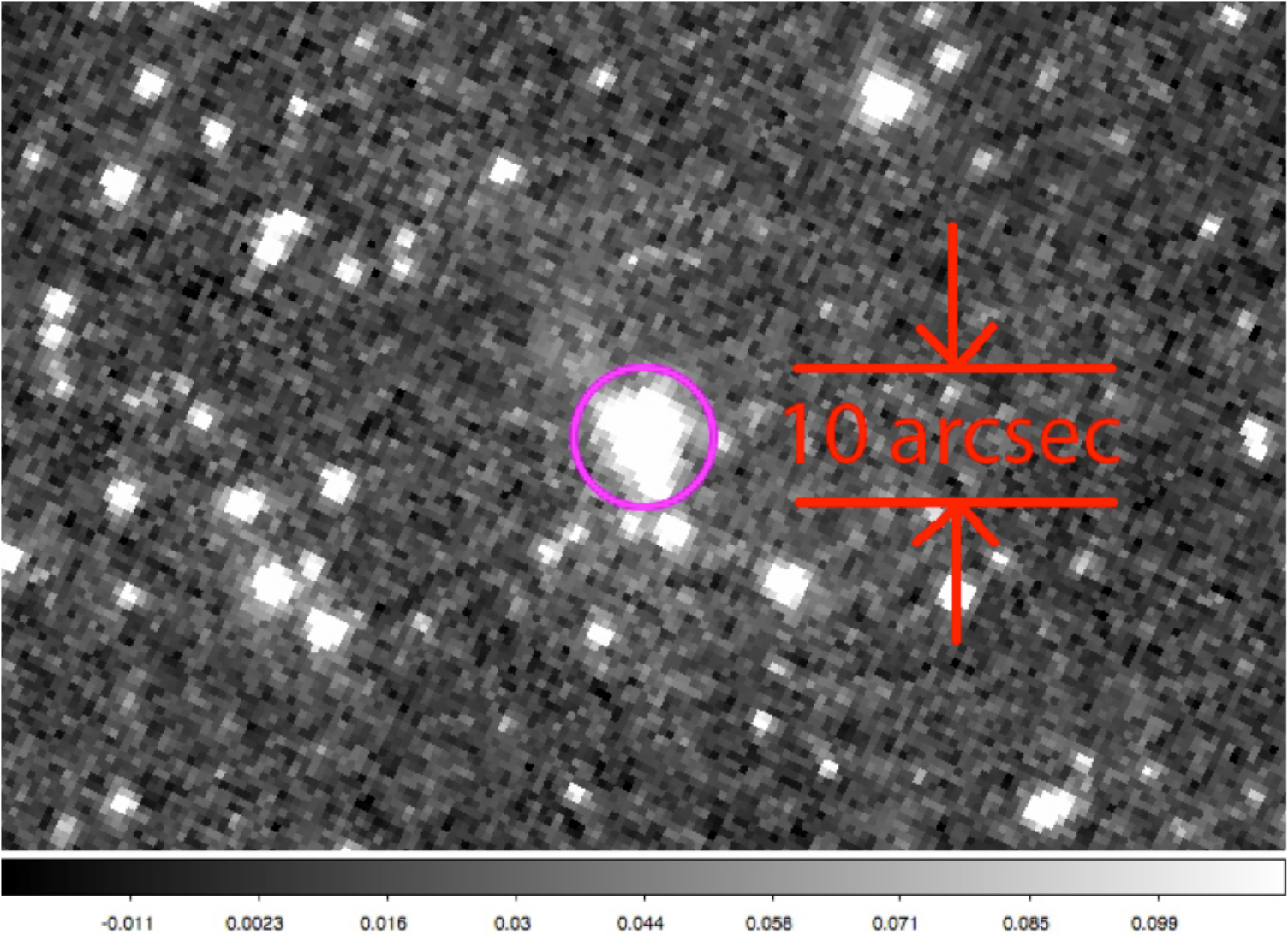}
\caption{Pic A W hotspot, \emph{Spitzer} 3.6 $\mu m$ image.}
\end{figure}

\clearpage
\begin{figure}
\includegraphics[width=0.99\textwidth]{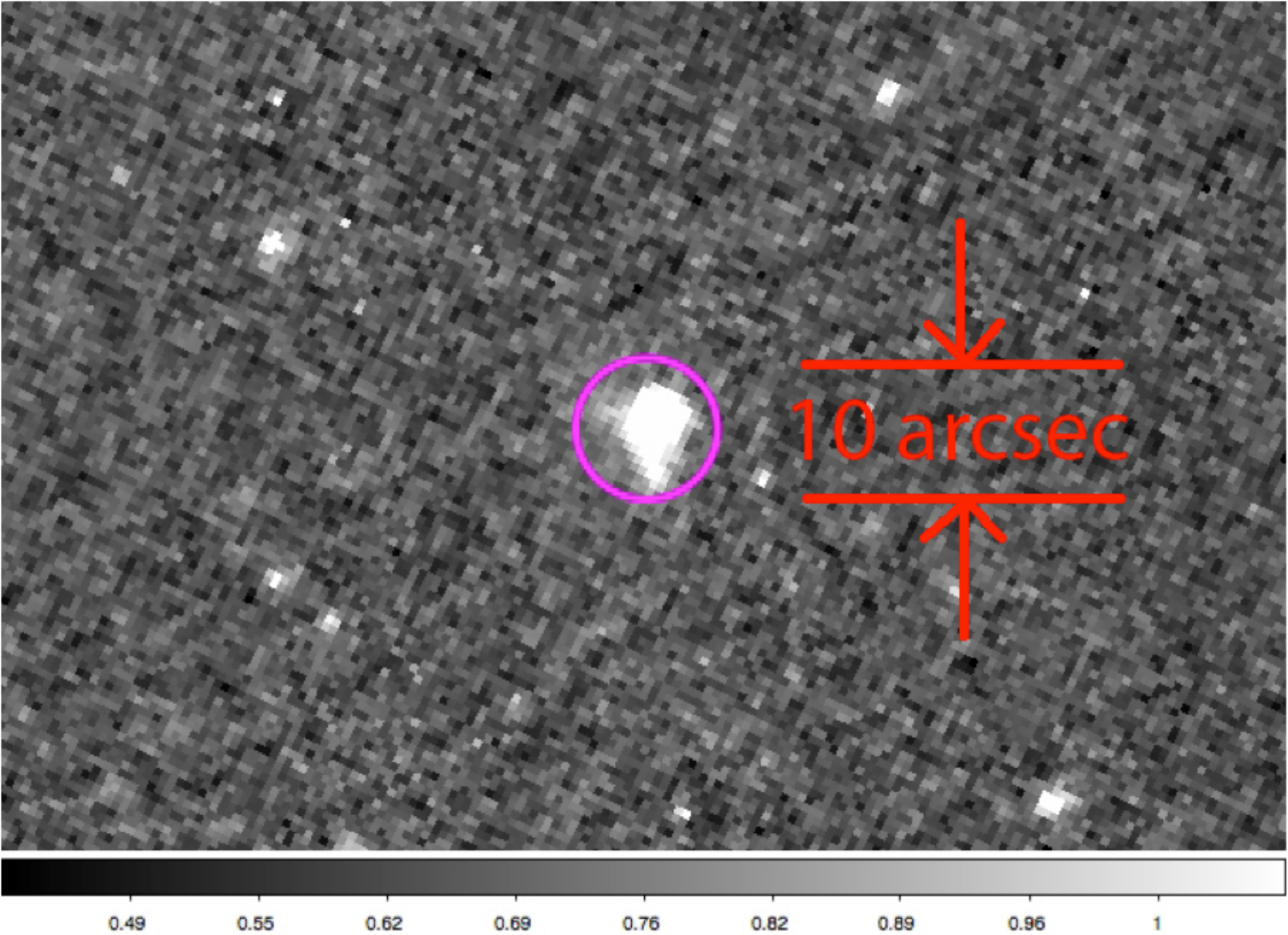}
\caption{Pic A W hotspot, \emph{Spitzer} 8 $\mu m$ image.}
\end{figure}

\clearpage
\begin{figure}
\includegraphics[width=0.99\textwidth]{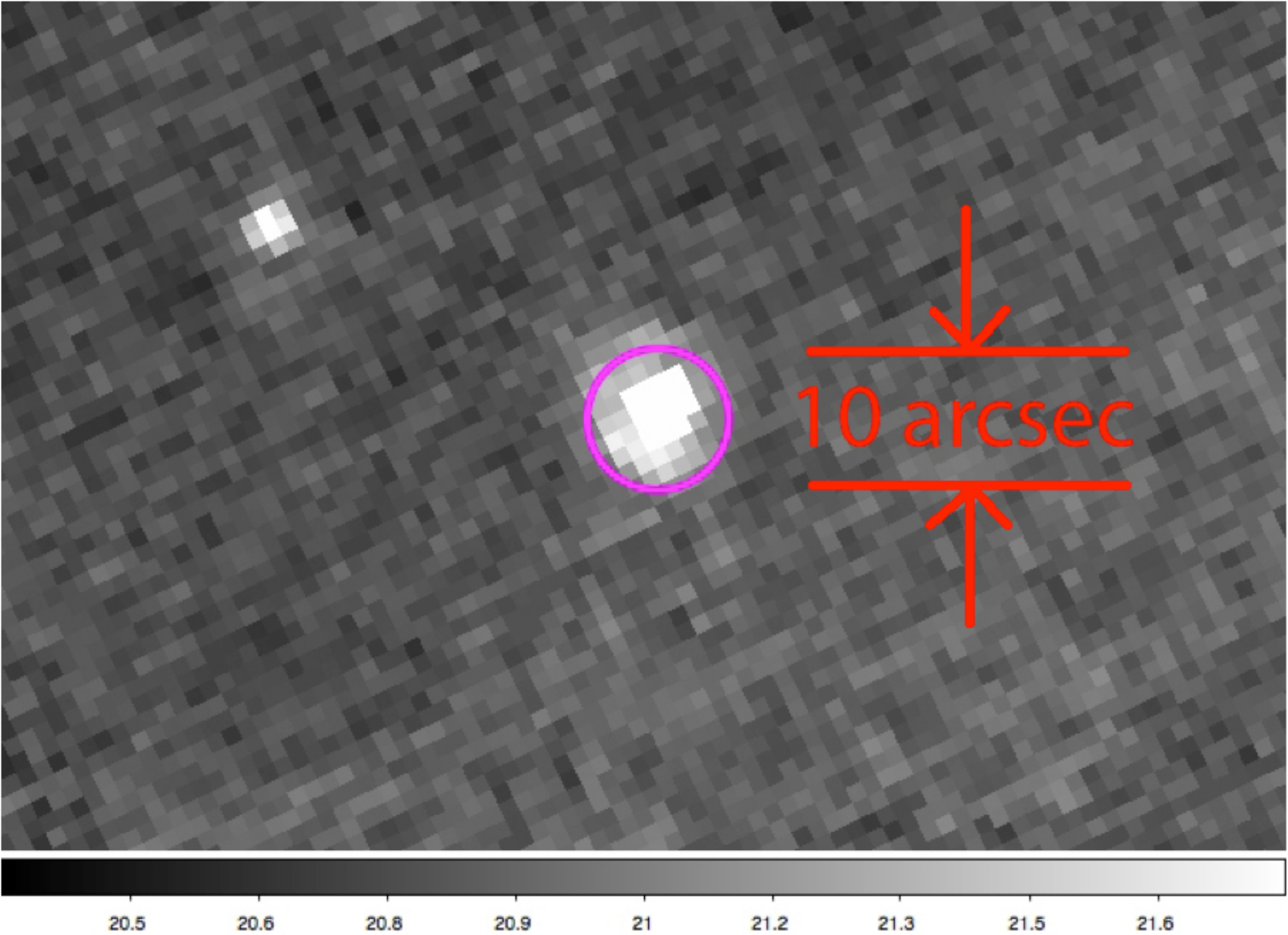}
\caption{Pic A W hotspot, \emph{Spitzer} 24 $\mu m$ image.}
\end{figure}

\clearpage
\begin{figure}
\includegraphics[width=0.99\textwidth]{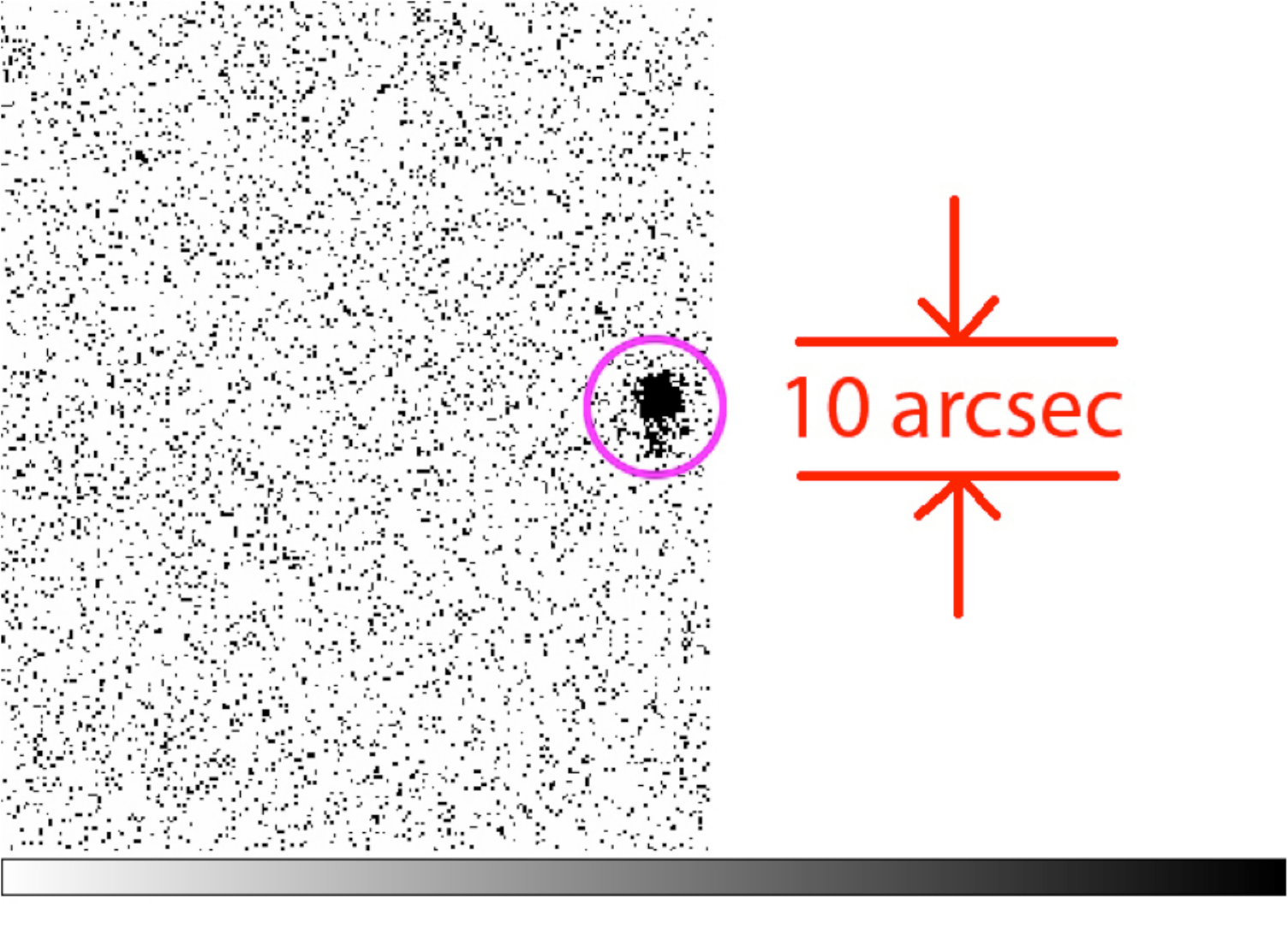}
\caption{Pic A W hotspot, \emph{Chandra} image.}
\end{figure}

\clearpage
\begin{figure}
\includegraphics[width=0.99\textwidth]{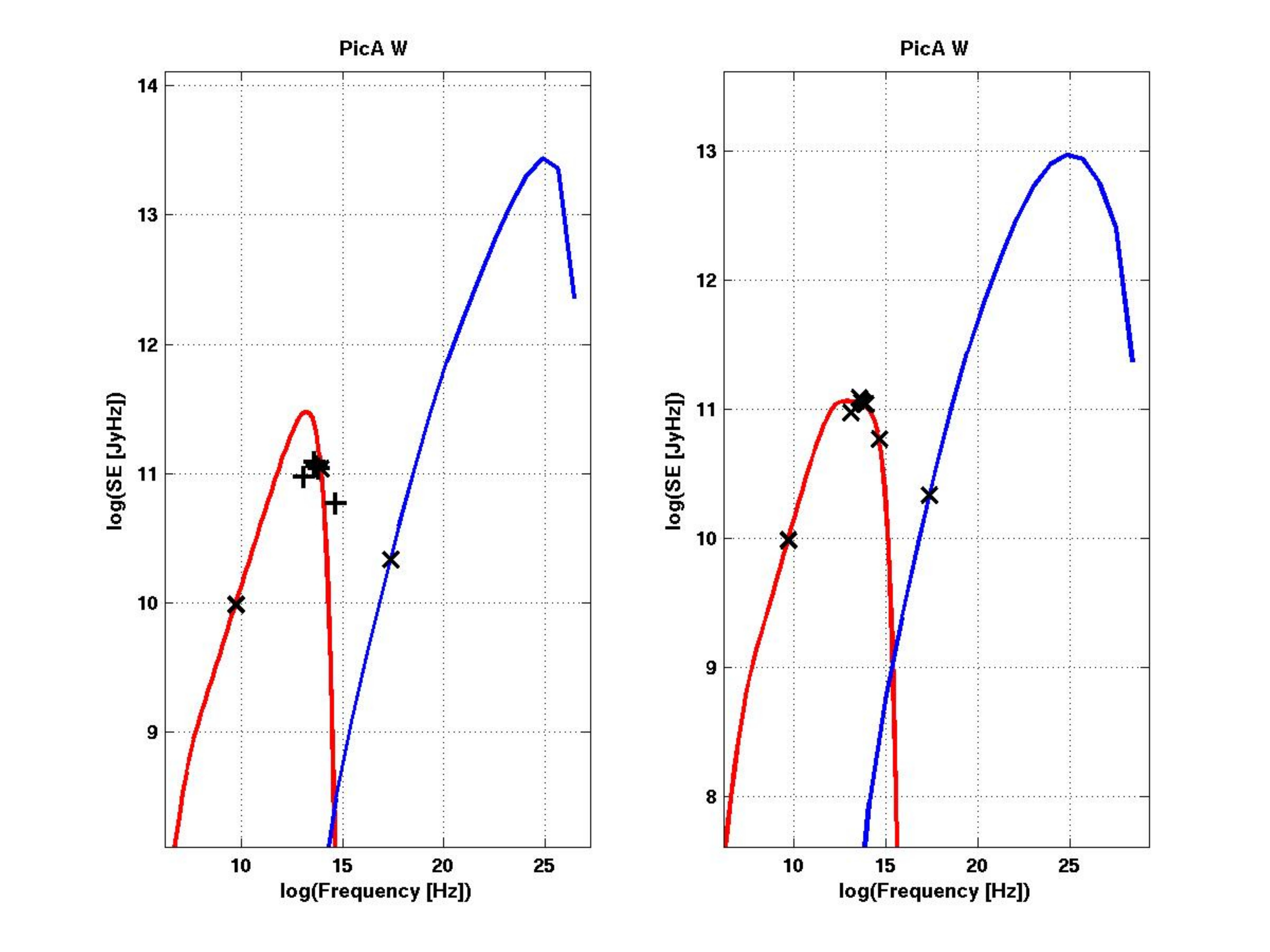}
\caption{Left: 3-fit model fitting of Pic A W. Right: all-fit model fitting of Pic A W. In red is the fitted synchrotron curve and in blue is the fitted IC curve. Cross symbols are used for SEs that are fitted; plus symbols are used for SEs not fitted.}
\end{figure}

\clearpage
\begin{figure}
\includegraphics[width=0.99\textwidth]{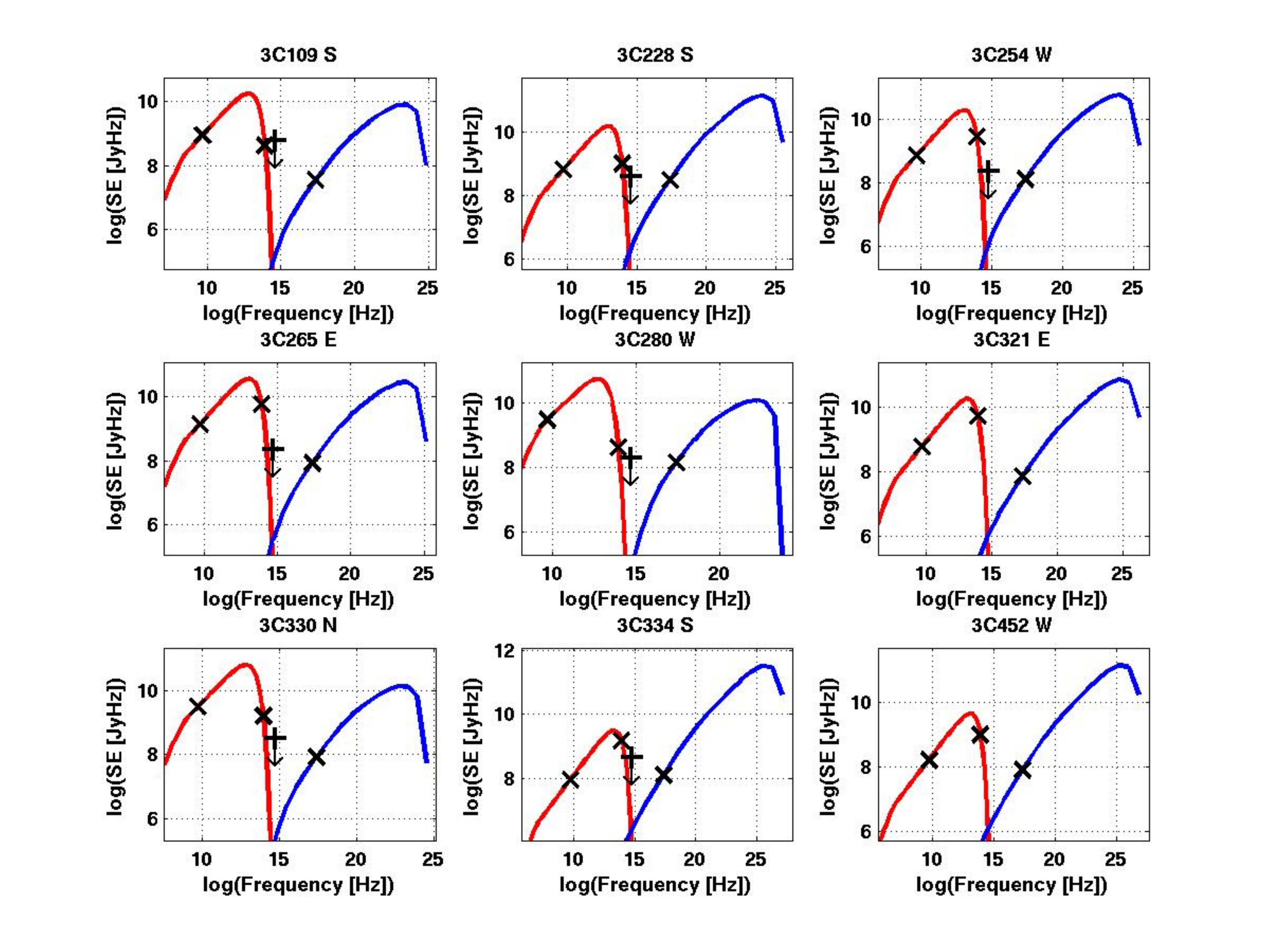}
\caption{SED plots for the 9 SWM hotspots obtained from the 3-fit model fitting. In red is the fitted synchrotron curve and in blue is the fitted IC curve. $SE_{R}$, $SE_{IR}$, and $SE_{X}$ are plotted as crosses and the optical upper limits that were not used in the fitting are plotted crosses with down arrows.}
\end{figure}

\clearpage
\begin{figure}
\includegraphics[width=0.99\textwidth]{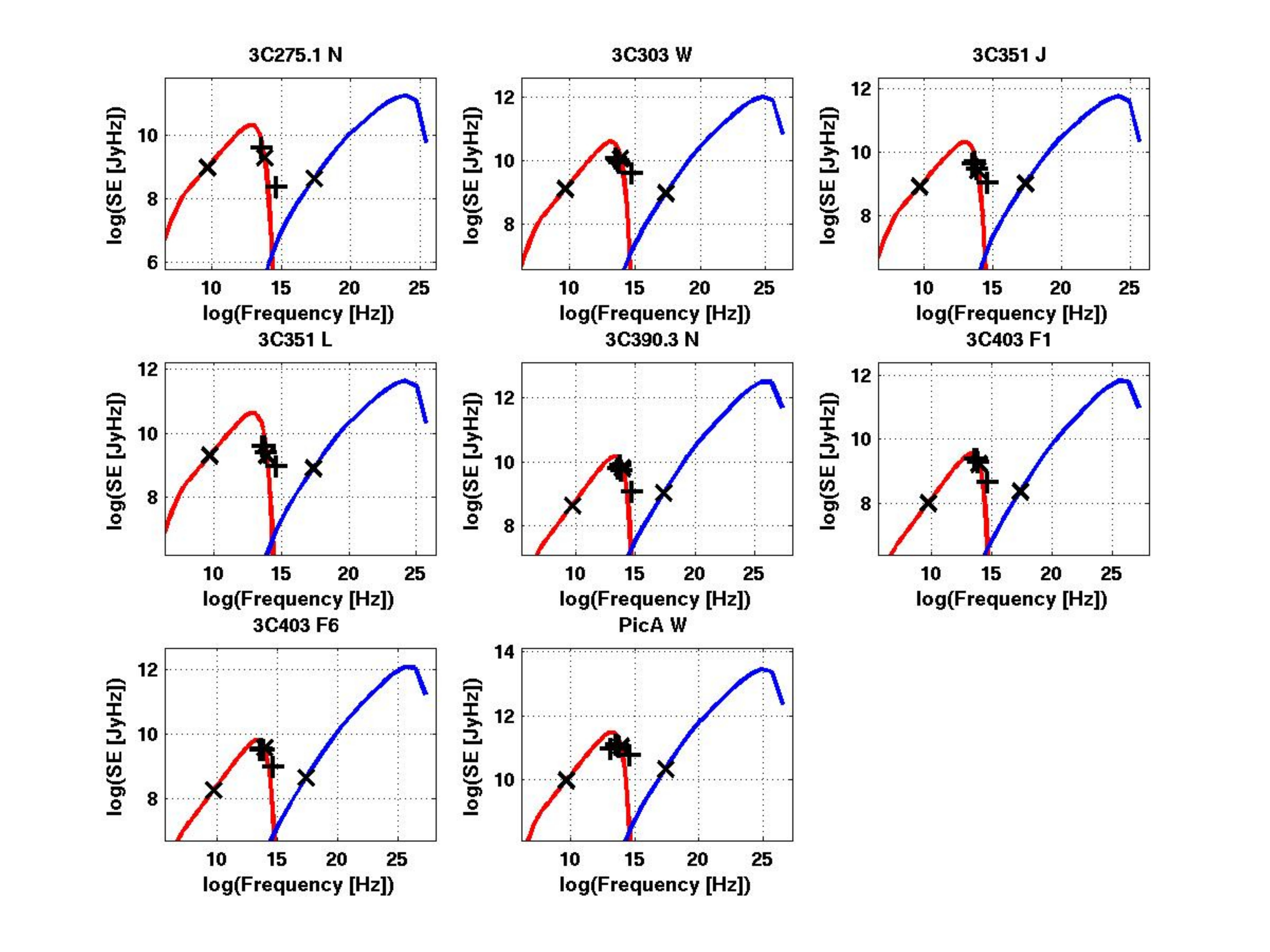}
\caption{SED plots for the 8 SCM hotspots obtained using the 3-fit model fitting. In red is the fitted synchrotron curve and in blue is the fitted IC curve. $SE_{R}$, $SE_{IR}$, and $SE_{X}$ are plotted with the x symbol and the other SEs not used in the fitting are plotted with the + symbol. Note that all optical SEs are measurements and not upper limits.}
\end{figure}

\clearpage
\begin{figure}
\includegraphics[width=0.99\textwidth]{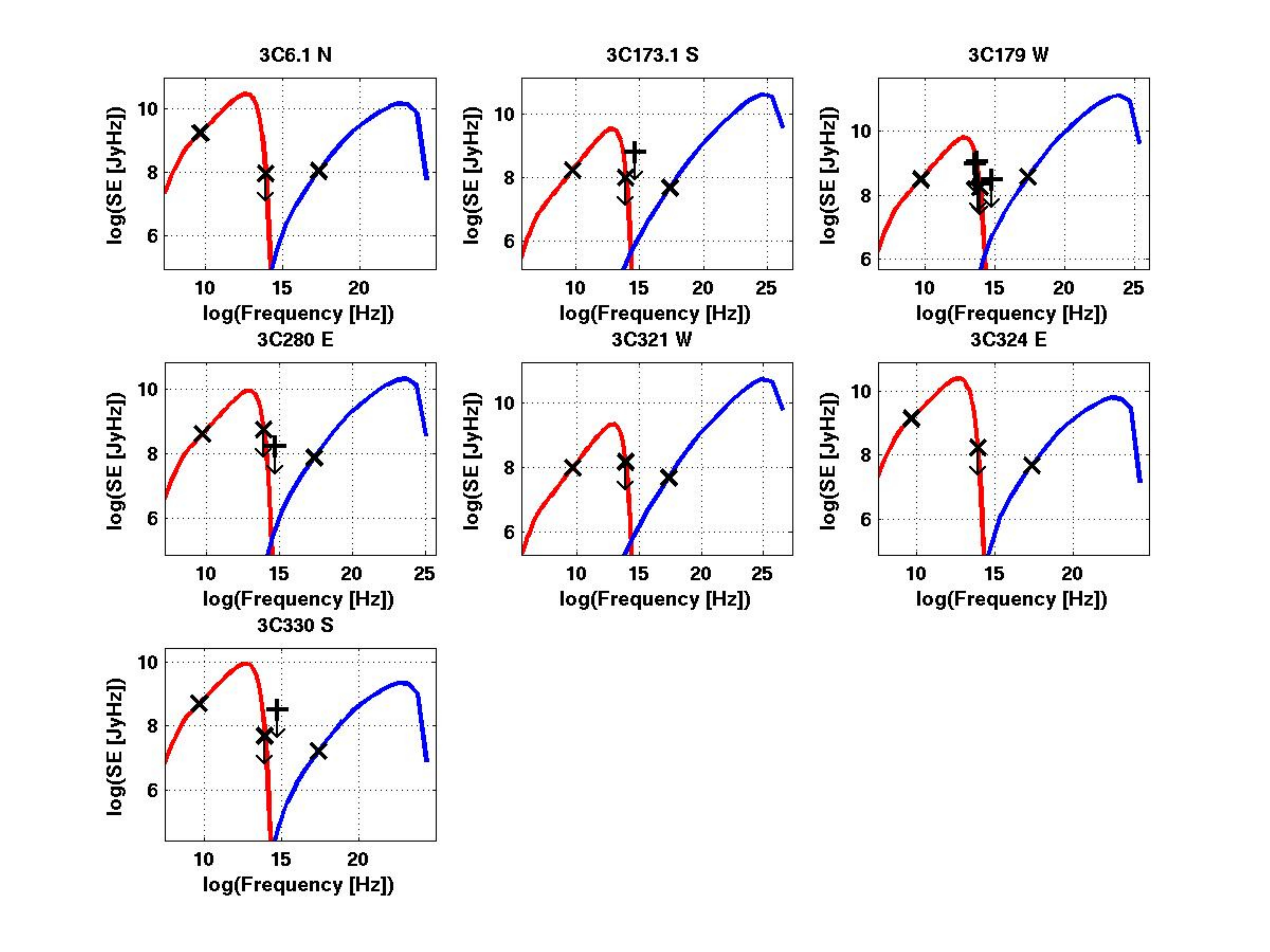}
\caption{SED plots obtained using the 3-fit model fitting for those hotspots in which there are only IR upper limits used for the $SE_{IR}$ values in this model fitting. In red is the fitted synchrotron curve and in blue is the fitted IC curve. $SE_{R}$, $SE_{IR}$, and $SE_{X}$ are plotted with the x symbol and the other SEs not used in the fitting are plotted with the + symbol. Upper limits are also plotted with a down arrow.}
\end{figure}

\clearpage
\begin{figure}
\includegraphics[width=0.99\textwidth]{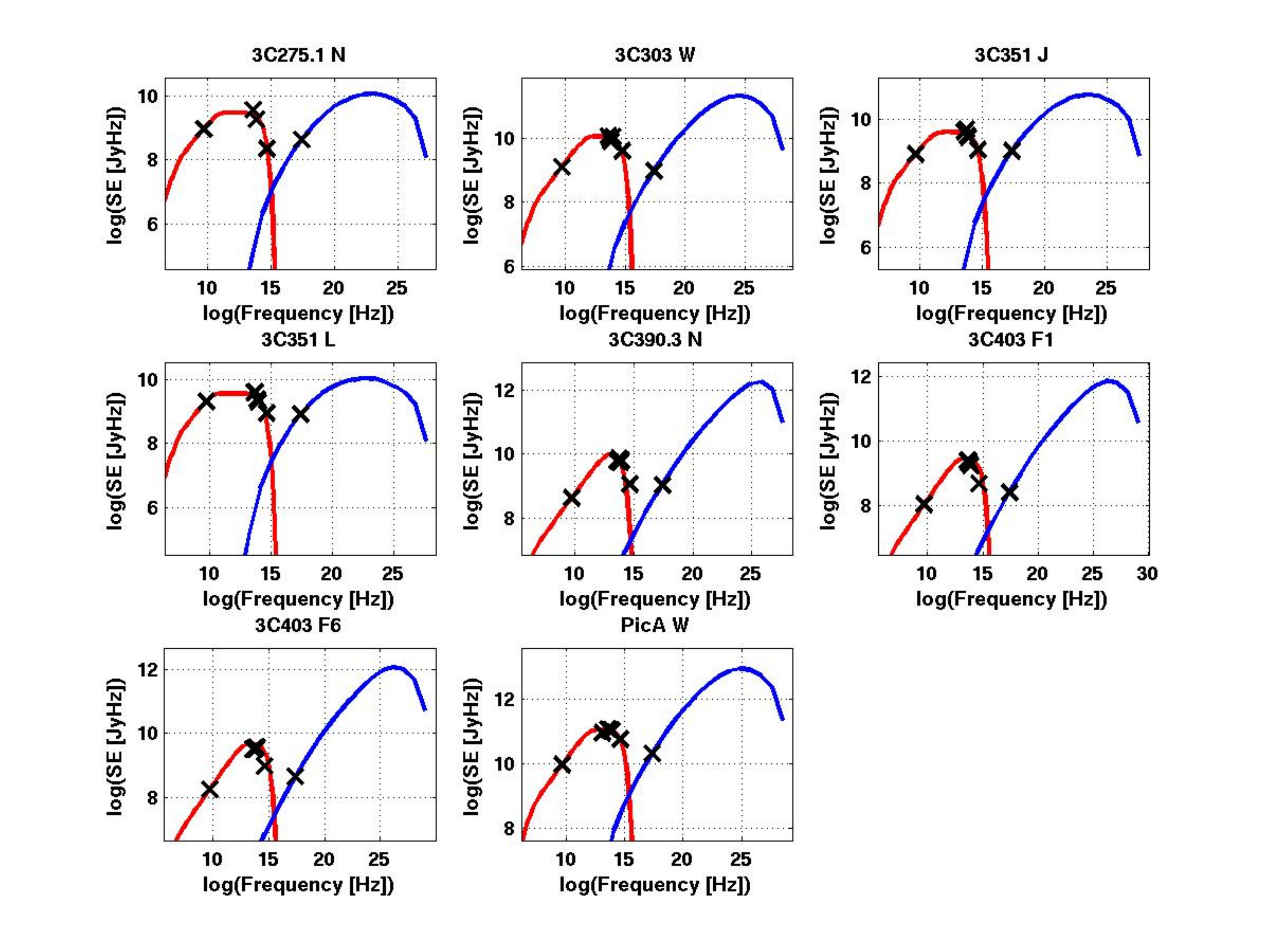}
\caption{SED plots for the 8 SCM hotspots obtained using the all-fit model fitting with measured SEs plotted as x symbols. In red is the fitted synchrotron curve and in blue is the fitted IC curve. Note that all optical SEs are measurements and not upper limits.}
\end{figure}

\clearpage
\begin{figure}
\includegraphics[width=0.99\textwidth]{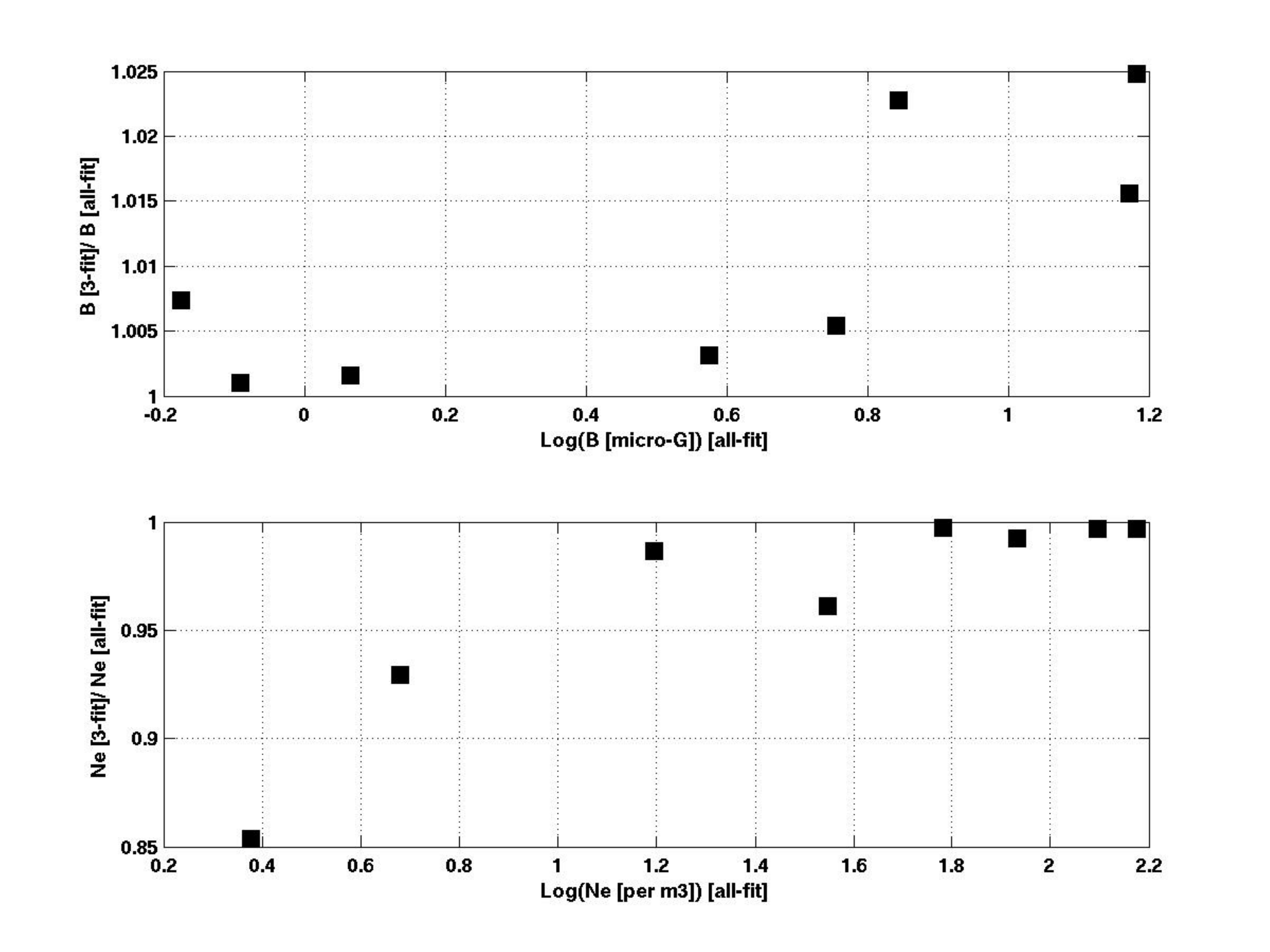}
\caption{Comparison of 3-flux density and all-flux density fitting the \emph{Spitzer} SCM hotspots. Top: Ratio of 3-fit B/all-fit magnetic field ($B$) versus log(all-fit $B$). Bottom: ratio of 3-fit and all-fit total electron density ($N_{e}$) versus log(all-fit $N_{e}$).}
\end{figure}

\clearpage
\begin{figure}
\includegraphics[width=0.99\textwidth]{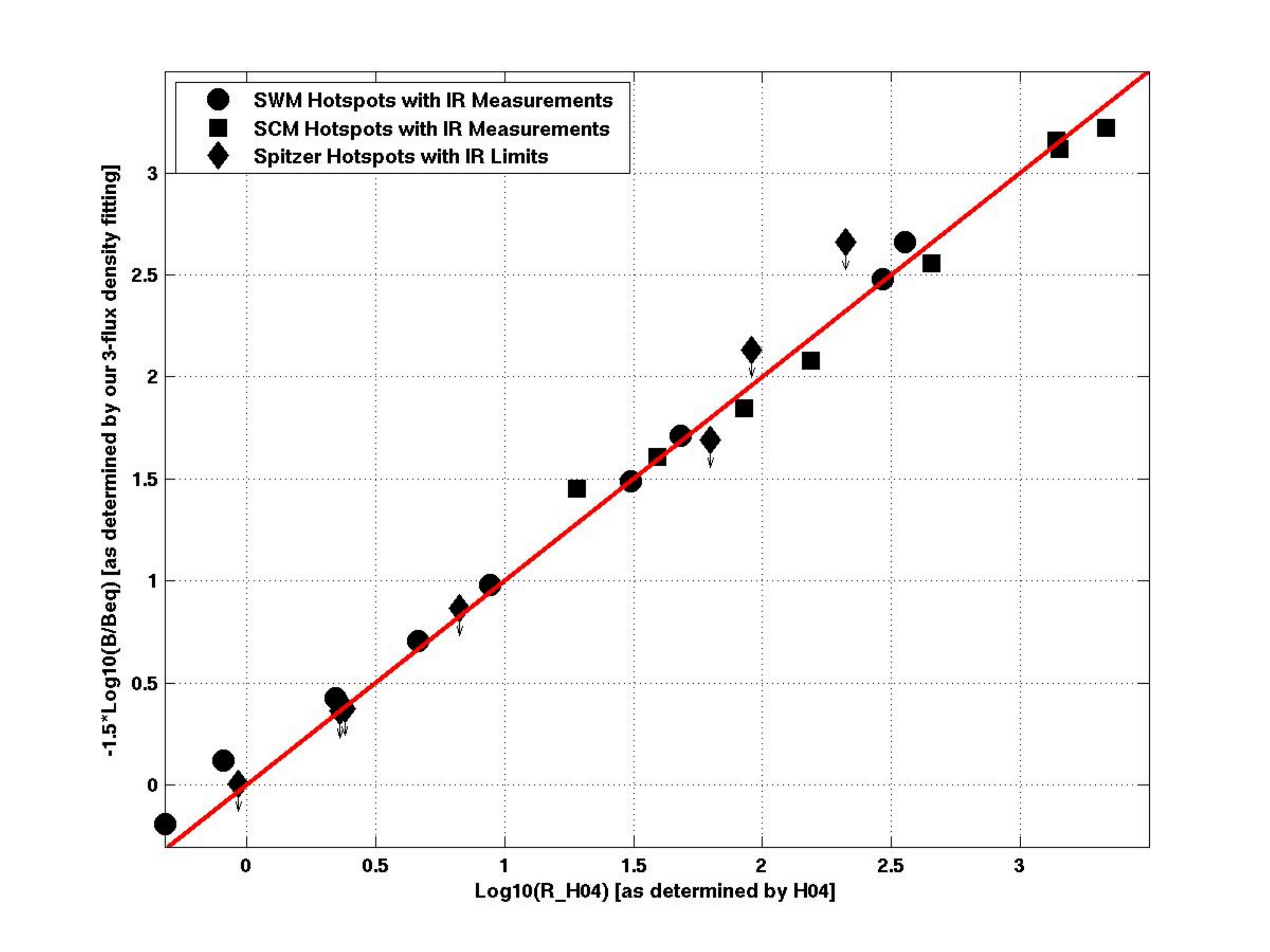}
\caption{Plot of $1.5log_{10}(B/B_{eq})$, as determined from our 3-fit model-fitting for all 24 hotspots in our sample using our SSC model (based on G09 SSC model), versus $R_{H04}$ as determined by H04 using their SSC model. Equation (5) shows that these should be the same quantity. For comparison, a line with unit slope is also plotted. Hotspots with IR upper limits are so indicated in this plot.}
\end{figure}

\clearpage
\begin{figure}
\includegraphics[width=0.99\textwidth]{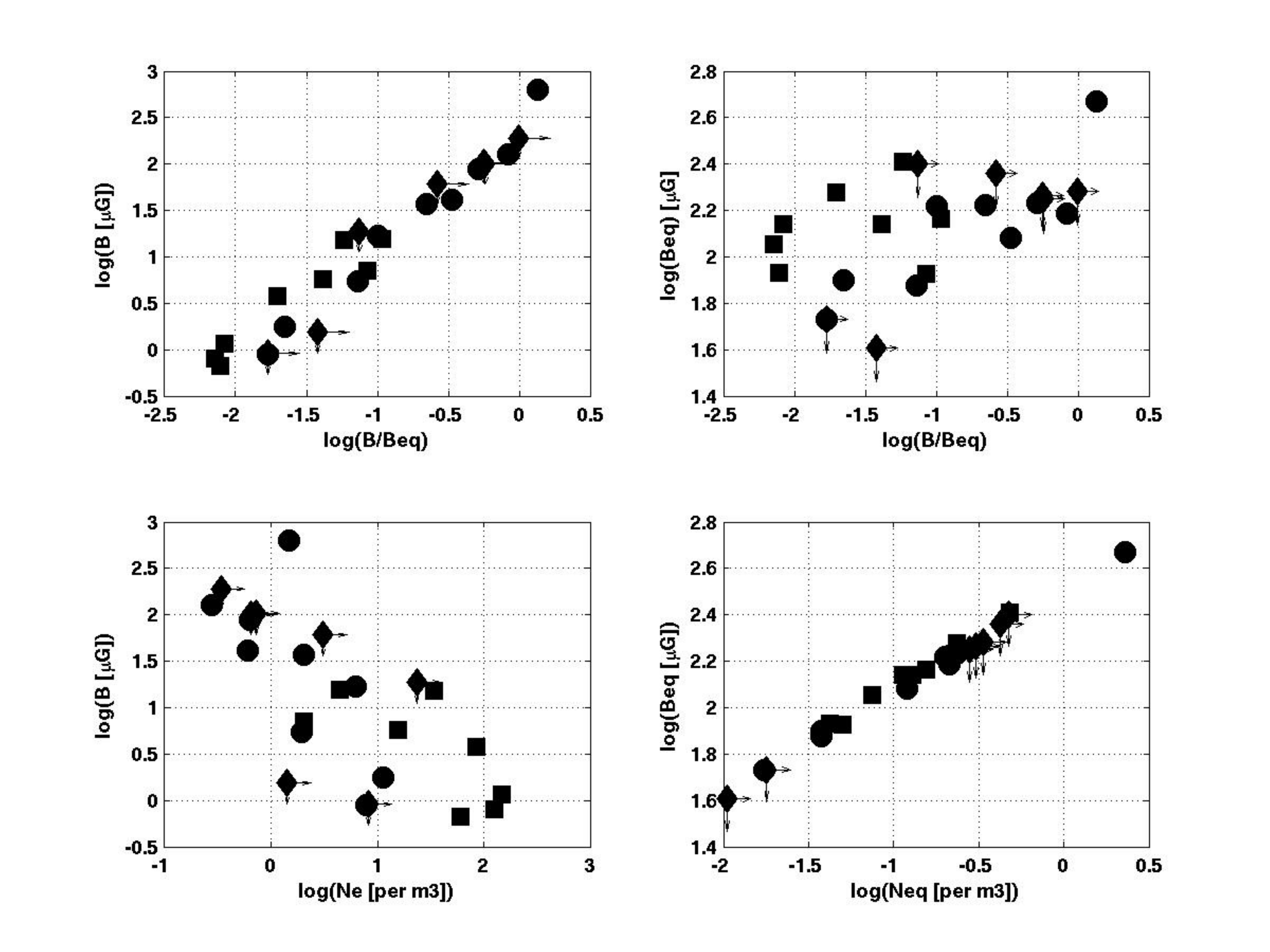}
\caption{3-fit model-fitting log-log plots of various quantities related to magnetic field and electron number density.  Top left: $B$ vs. $B/B_{eq}$. Top right: $B$ vs. $B/B_{eq}$. Bottom left: $B$ vs. $N_{e}$. Bottom right: $B_{eq}$ vs. $N_{eq}$. Circle, square, and diamonds symbols are for SWM hotspots with measured IR flux densities, SCM hotspots with measured IR flux densities and for hotspots with only upper limits respectively. Hotspots with IR limits are shown with arrows to indicate whether IR  upper limit provides an upper or lower limit on the quantity determined from fitting and plotted on each axis.}
\end{figure}

\clearpage
\begin{figure}
\includegraphics[width=0.99\textwidth]{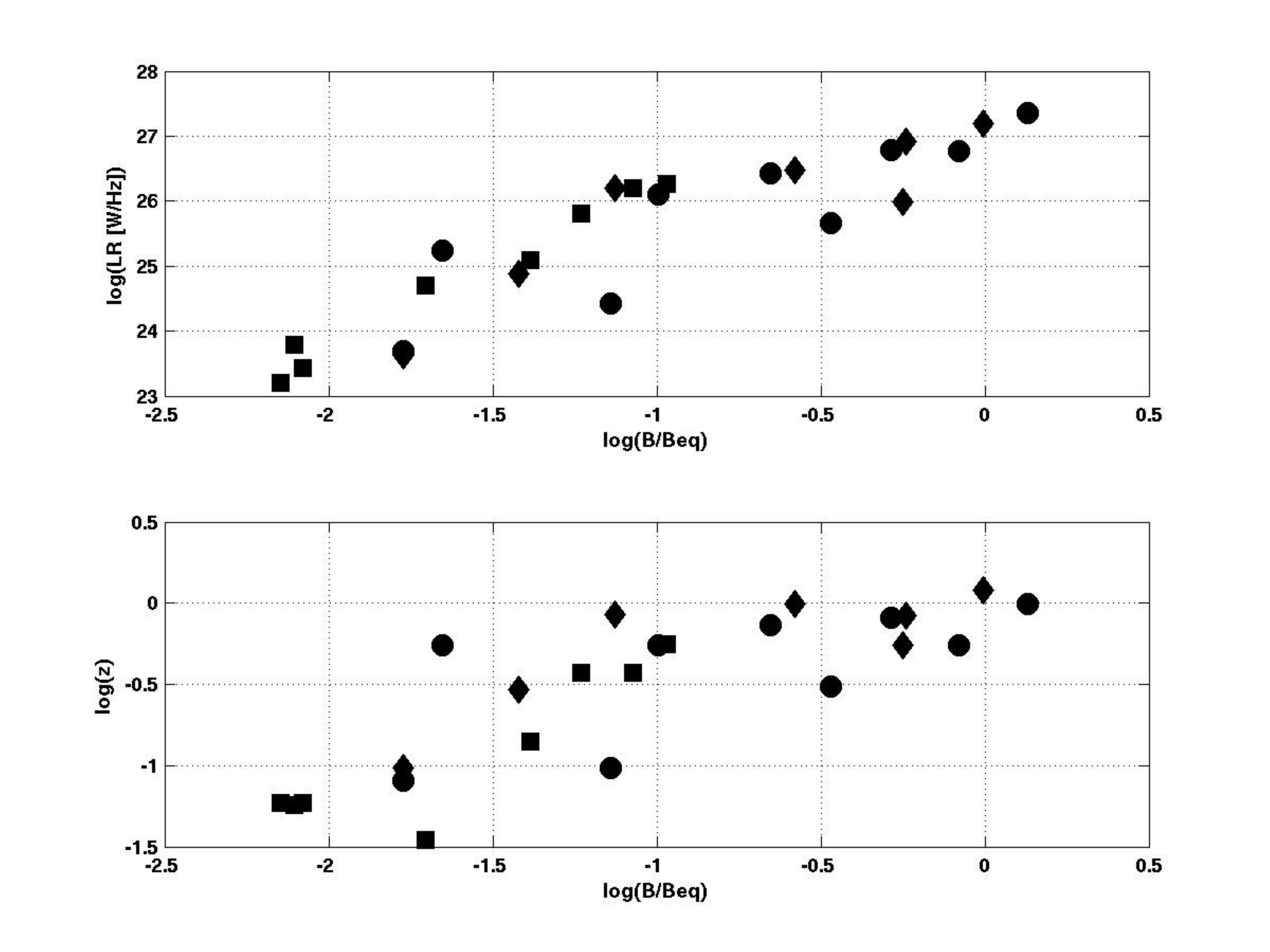}
\caption{3-fit model-fitting log-log plots of various quantities.  Top: $log(L_{R})$ vs. $log(B/B_{eq})$. Bottom: $log(z)$ vs. $log(B/B_{eq})$. Circle, square, and diamonds symbols are for SWM hotspots with measured IR flux densities, SCM hotspots with measured IR flux densities, and for hotspots with only upper limits respectively.}
\end{figure}

\clearpage
\begin{figure}
\includegraphics[width=0.99\textwidth]{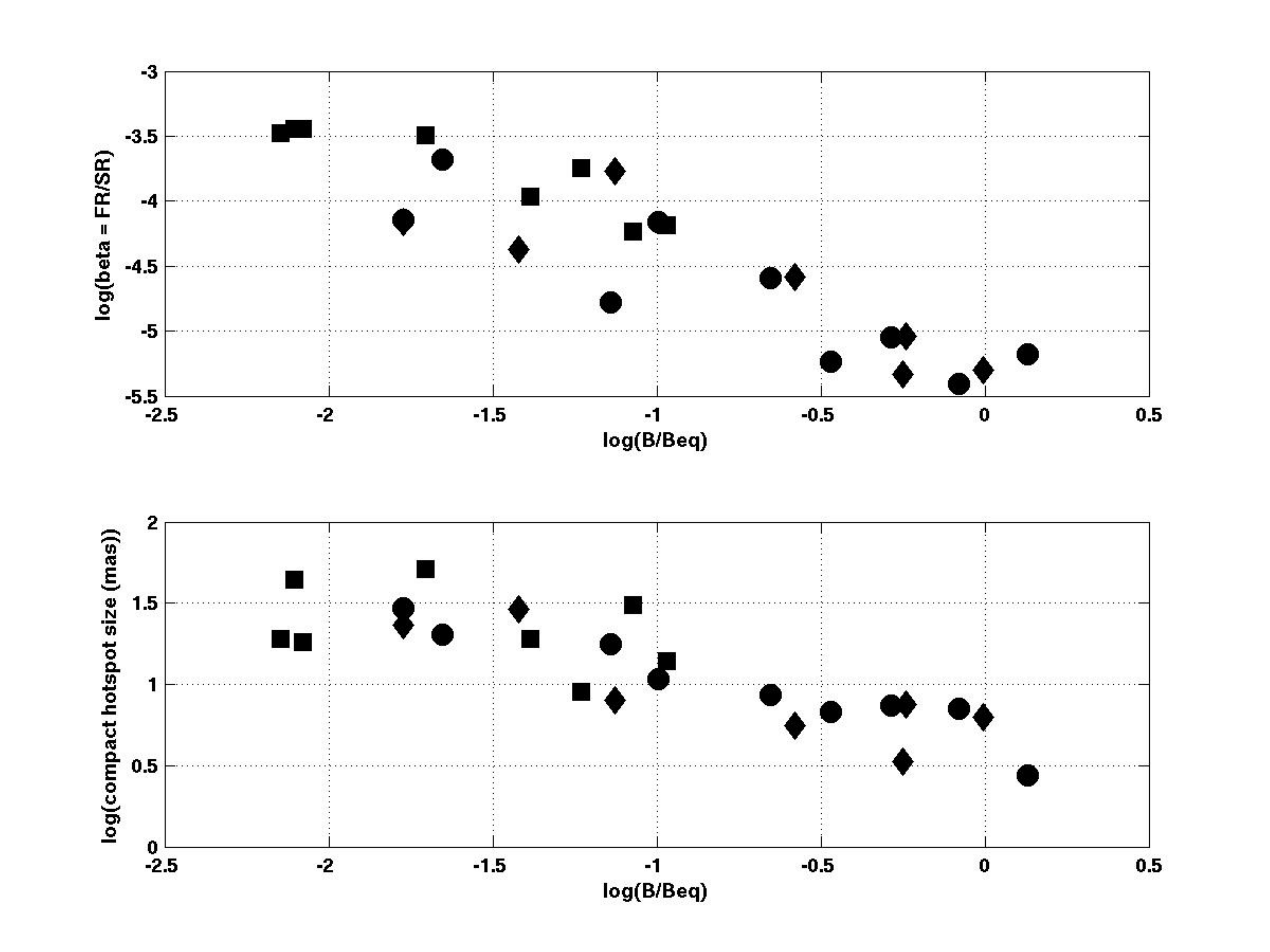}
\caption{Comparison of SO-CD and SSC model parameters. Top: SO-CD model $\beta$ parameter versus SSC $B/B_{eq}$ model parameter. Bottom: SO-CD model $\theta_{C}$ parameter versus SSC $B/B_{eq}$ parameter. Circle, square, and diamonds symbols are for SWM hotspots with measured IR flux densities, SCM hotspots with measured IR flux densities, and for hotspots with only upper limits respectively.}
\end{figure}

\end{document}